\pdfoutput=1
\documentclass[twocolumn,english,prb,showpacs,preprintnumbers,amsmath,amssymb,aps,10pt]{revtex4-1}
\usepackage[latin9]{inputenc}
\usepackage[english]{babel}
\usepackage{esint}
\usepackage{braket}
\usepackage{tikz}\usetikzlibrary{plotmarks}
\usepackage[colorlinks=true, urlcolor=blue, linkcolor=black, citecolor=black, filecolor=black, menucolor=black, pdfpagelayout=TwoColumnRight, pdfstartview=FitH, plainpages=false, pdfpagelabels,breaklinks=true]{hyperref}
\usepackage{enumitem}

\newcommand{\oncite}[1]{Ref. \onlinecite{#1} }
\newcommand{\oncitep}[1]{Ref. \onlinecite{#1}. }
\newcommand{\oncitec}[1]{Ref. \onlinecite{#1}, }

\newcommand{\ie} {{\em i.e. }}

\newcommand{\etal} {{\em et al. }}

\begin{document}
\title{ Effective Field Theory for a p-wave  Superconductor in the Subgap Regime}

\author{T.H. Hansson}
\author{T. Kvorning}
\affiliation{Department of Physics, Stockholm University, AlbaNova University
Center, SE-106 91 Stockholm, Sweden}

\author{V.P. Nair}
\affiliation{Physics Department, City College of the CUNY, New York, NY 10031}

\author{G. J. Sreejith}
\affiliation{NORDITA, KTH and Stockholm University, Stockholm, Sweden }
\affiliation{MPI-PKS, Dresden, Germany}

\begin{abstract}
We construct an effective field theory for the 2d spineless p-wave paired superconductor that faithfully describes the topological properties of the bulk state, and also provides a model for the subgap states at vortex cores and edges. In particular it captures the topologically protected zero modes and has the correct ground state degeneracy on the torus. We also show that our effective field theory becomes a topological field theory in a well defined scaling limit and that the vortices have the expected non-abelian braiding statistics.

\end{abstract}
\maketitle

\section{Introduction}
In topological phases of matter, such as quantum Hall (QH) liquids, superconductors, topological insulators and spin liquids,\cite{fradkin13,*bernevig13} the excitations in the interior of the system are separated from the ground state by an energy gap, thus distinguishing them from ordinary metals or magnets. They however differ in important ways from trivial gapped phases, such as conventional band insulators, typically by having protected gapless edge modes. Particularly fascinating are the topologically ordered states characterized by long range gauge interactions and fractionalized quasiparticles in the bulk.

An important theoretical approach to  topologically ordered phases is based on topological field theories (TFT), which directly builds in important features such as the absence of bulk excitations, ground state degeneracy on topologically non-trivial manifolds, and dynamical edge states.  In the case of two space dimensions, the TFTs can also capture the fractional  braiding statistics of the quasiparticles that can be abelian or non-abelian.

The prototypical TFTs pertinent to topologically ordered phases, are the multi-component abelian Chern-Simons (CS) theories, proposed by Wen, (see  \emph{e.g.} Ref. \onlinecite{wen95}) to describe the hierarchical QH liquids. It was also early realized by Wen that ordinary s-wave superconductors are topologically ordered, and for the two-dimensional case the corresponding TFT, which is of BF type, was formulated and analyzed in \oncitep{hos} This BF theory, which describes a time-reversal invariant (TRI) state, is very closely related to the CS theory for the QH liquids. Simply put, it has two components corresponding to the two chiralities needed to retain TRI. The BF theories can, however, be generalized to higher dimensions also, and the one describing the 3d s-wave superconductor was originally constructed by Balachandran \etal\cite{balachandran93}.   Cho and Moore\cite{cho11}, and more recently Chan \emph{et al.}\cite{atma}, used the BF theory to describe 3d topological insulators, and an interesting feature of their construction is that a purely bosonic TFT describes fermionic edge modes.

All the examples discussed above are abelian, and it has turned out to be more difficult to construct TFTs for non-abelian topological phases. The most thoroughly analyzed non-abelian state is the Moore-Read (MR) pfaffian state which is likely to describe the observed QH liquid at filling fraction $\nu = 5/2$.\cite{mr} The MR can be understood as a paired state of composite fermions\cite{greiter92}, and  the precise connection to a spin-less $p_x + ip_y$ superconductor was made by Read and Green\cite{read00}. The aim of this paper is to construct and analyze a low energy theory for this topological superconductor.
 
An important property of the MR state, which it shares with the spinless 2d $p_{x}+ip_{y}$ superconductor\cite{read00}, is that the fundamental vortices support localized, and topologically protected, zero energy Majorana modes. As a consequence, a set of $2n$ vortices at fixed positions define a Hilbert space of dimension $2^{n-1}$, and braiding the vortices corresponds to unitary rotations in this space. This is the basis of the non-abelian fractional statistics that has been looked for in experiments\cite{willett} and is proposed to be useful in quantum information applications\cite{qinfo}. For a system with a boundary, there are chiral, low energy, edge modes, which are strictly gapless if the total vorticity of the bulk is odd\cite{fendley07}. The non-abelian nature of the state is also reflected in the ground state degeneracy on nontrivial manifolds. Specifically, the degeneracy on a torus is three, while it is four for an s-wave paired state\cite{read96}.
 
The properties just mentioned have mainly been deduced from the BdG description of the p-wave paired state. A self-consistent calculation in the presence of vortices and edges is only possible using numerical methods, but the existence and topological properties of localized solutions were established assuming various fixed mean field configurations \cite{read00,fendley07}. Later an index theorem argument for the existence of zero modes in odd vortex backgrounds was given in  \oncitep{tewari07}. 
 
Just as in the case of the abelian quantum Hall effect, it would be of interest to have an effective low energy theory of the p-wave superconductor that encodes both the topological information about quasiparticles and vortices, and the dynamics of the fermionic subgap modes at vortex cores and at edges. In the quantum Hall case the relevant microscopic theory which is amenable to mean field approximations, is the Ginzburg-Landau-Chern-Simions  (GLCS) theory, and the effective low energy theory is the CS theory referred to earlier. This is a TFT, but by adding higher derivative Maxwell terms yields a dynamical Chern-Simons-Maxwell(CSM) theory that  also captures scale dependent quantities, such as the energy of low-lying gapped states and the size of the quasiparticles\cite{zhang92}. In simple cases, such as the Laughlin states, the CSM theory can be derived from the GLCS theory, but for general states this has so far not been possible. 

As already  mentioned, the TFT pertinent to the s-wave superconductor is a BF theory, and by again  adding Maxwell terms one can also describe features such as the magnetic screening length and the frequency of the plasmons. In this paper we construct the corresponding theory for the case of p-wave pairing. 

Our starting point is  the TFT for the p-wave superconductor  proposed in  \oncite{hks} where the BF theory for the s-wave superconductor, based on two gauge fields $a$ and $b$, was  augmented with a fundamental Majaorana  field $\gamma$. Being neutral, the Majorana field, $\gamma$ was not coupled to the gauge field by a minimal coupling but by a topological Pauli like term $\sim da \gamma d\gamma$. This makes $\gamma$ dynamical only where the field strength $da$ is non-vanishing, \ie at vortex cores and at edges. Being a purely topological theory, the vortices do not have any extent, and the related fermionic zero modes are also pointlike. This singular nature of the vortex modes causes problems, in particular there are zero mode solutions both for even and odd vortices, and similarly gapless edge modes irrespective of the vorticity of the bulk\cite{hks}. 

Naively one might think that these deficiencies could be remedied by adding Maxwell terms to regularize the vortex cores, just as in the s-wave case. This, as will be explained later, is not the case, and to regularize the ``$\gamma BF$'' theory of \oncite{hks} one needs a more sophisticated treatment of the vortex cores that properly describes the localized Caroli - de Gennes - Matricon modes below the fermi level.  

In this paper we propose the  ``$\psi$BF theory'',  based on two electrically neutral fermi fields, 
 as a model for the two-dimensional spinless p-wave superconductor. This theory is appropriate for the extreme type II case, and in the limit of zero charge screening length, \ie both the correlation length and the 
charge screening length  are put to zero while the London length $\lambda_L$ is kept finite. The theory is not topological, but in a scaling limit, that will be precisely defined, it only retains topological information. In addition to correctly encoding the topological features of the superconductor, it also provides a model for the subgap states that is consistent with previous numerical results using the BdG equations.

In the next section we first review the BF-Maxwell (BFM) description of superconductors, and then explain the extension needed to describe the limit of exact local charge screening. In section \ref{section:psiBF} we review the topological model in \oncitec{hks} and its limitations, and in the following section, which is the most important part of the paper, we define and discuss the  $\psi$BF model. In section \ref{section:subgap}  we 
give analytical solutions for the zero modes at vortices, as well as numerical solutions for the subgap spectra for states localized at both vortices and edges. In this connection we also present a specific model for the edge of the p-wave superconductor. In section \ref{section:GS-degeneracy} we determine the ground state degeneracy on a torus, with some technical details deferred to an appendix. Section \ref{section:topologicalscaling} defines the topological scaling limit, and in section \ref{section:non-abelian} we give a detailed derivation of the non-abelian statistics, again with some details  in an Appendix. The last section discusses the relation to previous work on effective topological theories for the Moore-Read quantum hole state and offers some speculations about possible extensions of the present work. An early, unpublished, version of this work\cite{old} employed a pairing term that supported static vortex solutions, but did not give the right  ground state degeneracy on the torus, and also involved a somewhat artificial boundary condition.

\section{The gauge field lagrangian}
\subsection{General discussion}
We start from a topological description of superconductors in terms 
of BF gauge theory which is reviewed in \oncitep{hos} In this theory, the quasiparticle 
current $j_q$ couples to a gauge field $a$, and in the 2+1 dimensional case which we will concentrate on in the following,  the 
point like vortices are are described by the current  $j_v$  that couples to the gauge field $b$, 
\begin{align}
{\mathcal L}_{BF}= & \frac{1}{\pi}\epsilon^{\mu\nu\rho}\partial_{\mu}a_{\nu}b_{\rho} -j_{{\rm q}}^{\mu}a_{\mu}-j_{{\rm v}}^{\mu}b_{\mu} \, .\label{bflag}
\end{align}
In 3+1 dimensions, $b$ is an antisymmetric tensor field (or a two-form) which couples to the world sheet of the propagating vortex string. 
In addition to the two local  gauge symmetries, this TFT, \eqref{bflag}  is also invariant under parity transformations $(x,y)\to(-x,y)$ where the fields transform as $(a_0,a_x,a_y)\to(a_0,-a_x,a_y)$ and $(b_0,b_x,b_y)\to(-b_0,b_x,-b_y)$; and under time reversal where the fields transform as $(a_0,a_x,a_y)\to(a_0,-a_x,-a_y)$ and $(b_0,b_x,b_y)\to(-b_0,b_x,b_y)$.

It is known that by supplementing a TFT with non-topological terms more of the low energy physics can be described.  In the context of an effective local field theory it is natural to extend \eqref{bflag} by adding Maxwell terms, which will turn point like charges and vortices into  exponentially localized charge and vorticity distributions. 
Although in a real superconducting film coupled to 3+1 dimensional electromagnetism there are  long range interactions that give screening by a power laws\cite{hos}, we shall here for simplicity model extended vortices by including the 2+1 dimensional Maxwell terms, 
\begin{align}
\label{max}
\mathcal{L}_{M}&=  \frac {\alpha_1} {4\pi} (\vec {E}^{a})^{2}-\frac{\alpha_{2}}{4\pi}(B^{a})^2 +\frac{\beta_{1}}{4\pi}(\vec{E}^{b})^{2}-\frac{\beta_{2}}{4\pi}(B^{b})^{2}
\end{align}
where $B^{b}=\epsilon^{ij}\partial_{i}b_{j}$ \emph{etc.,} and we define the the BF-Maxwell (BFM) theory  by
\begin{align}
{\mathcal L}_{BFM}  = {\mathcal L}_{BF} + {\mathcal L}_M \label{bfmlag} \, .
\end{align}
In the  pure BF-theory \eqref{bflag}, pointlike currents $j_q$ and $j_v$, yield pointlike gauge field configurations by relations like
$ \pi \rho_v= B^{(a)}$,  $ \pi \rho_q= B^{(b)}$ {\em etc.}. Although the charges are completely screened, there is still a long range interaction giving rise to the mutual braiding phase factor -1.

To make a more detailed model for the charge and vorticity distributions we could add higher order terms in the field strengths, but such microscopic details are of no importance for the following, so we will stick to the simplest choice which is to use the quadratic Maxwell terms.

The BFM-theory \eqref{bfmlag}  describes vortices and charges with  spatial extents $\lambda_{L}=\sqrt{\alpha_{2}\beta_{1}}$, and $\lambda_{D}=\sqrt{\alpha_{1}\beta_{2}}$, and their braiding phase is  well-defined only when charges and vortices are separated by distances much larger than the screening length. In addition to these length scales the BFM theory also encodes the plasma frequency $\omega_{p}^{-1}=\sqrt{\alpha_{1}\beta_{1}}$ and the vortex energy $\epsilon_{v}=1/\alpha_{1}$. 

It is important to realize what kind of superconductor that can  be described by the BFM-theory. 
From the solution for a vortex given in section \ref{section:vortedge} below, it follows that  the
current density never go to zero at the center of the vortex, which translates into having zero correlation length corresponding to the extreme type II limit. The BFM theory does not provide any microscopic description of the low-lying fermion modes that are localized at vortex cores and at edges. In spite of this,  it gives a consistent description of an s-wave superconductor at energies below the lowest of the minigap $\Delta_m \sim \Delta^2/k_F$ for the vortex core states, and the edge gap $\Delta_{ed}$ which
is generically opened due to back scattering at the edge. 
In the p-wave case, on the contrary, the BFM theory is not correct at any energy scale due to the topologically protected fermionic zero modes.

\subsection{\texorpdfstring{The  Lagrangian ${\mathcal L}_g$}{The Lagrangian for the guage fields}}
To proceed, we recall how a vortex is described in the BdG formalism. Starting from some vortex background in the order parameter field $\Delta(\vec r. t)$, one can find the localized subgap modes, as well as the continuous spectrum. In the simplest approximation, one then just 
fill up the localized modes below the fermi level, and in the s-wave case there is a minigap to the lowest vortex state. This vortex configuration is not electrically neutral since the depletion of the order parameter is not fully compensated by the localized fermion modes.
In a more refined calculation where the full fermion spectrum consisting of both the distorted plane waves, and the localized modes, is self-consistently determined, the vortex is strictly neutral. Also, in a real superconductor, the charge screening length is typically much smaller then both the correlation length and the London length.

With these facts in mind, we shall construct an effective theory for a superconducting state where the charge screening is local, and  where consequently  the subgap modes are described by electrically neutral fermions. The most obvious way to achieve the first objective would seem to be to use the BFM lagrangian \eqref{bfmlag} and take $\lambda_D=\sqrt{\alpha_1\beta_2}\rightarrow 0 $, while keeping $\lambda_L$  fixed. Although this is logically an option, we have not found any  consistent way to incorporate the subgap modes. 
Instead we shall use a simpler formulation where exact local charge screening is built in from the start, while the vortex size $\lambda_L$ is kept finite. As a consequence, the fermions, both the usual quasiparticles above the superconducting gap which are described by $j_q$, and the subgap modes to be discussed below, are strictly neutral with respect ot the electromagnetic field $a$ and couple only to a statistical potential $\omega$. In the BF limit, the fields $a$ and $\omega$ are identical, but when the Maxwell terms are added, they have to be distinguished. As we shall demonstrate, this is  achieved by using the Lagrangian,
\begin{align} \label{gaugelag}
{\mathcal L}_g = \frac 1 \pi adb + {\mathcal L}_M +
 \frac 1 \pi d\omega (\tilde b - b) - j_v \tilde b - j_q \omega  \, .
\end{align} 
First consider the case where ${\mathcal L}_M $ is absent. Then we can integrate $b$ to get 
${\mathcal L}_g = \frac 1 \pi \omega d\tilde b  - j_v \tilde b - j_q \omega $, which is just \eqref{gaugelag} after the  
identifications $\omega\rightarrow a$ and $\tilde b \rightarrow b$. With $\mathcal L_M $ present,  
 $\tilde b$ is a multiplier field that determines the singular potential $\omega$ in terms of the vortex sources as
\begin{align} \label{spot}
d\omega = \pi j_v \, ,
\end{align}
which when substituted into \eqref{gaugelag} gives, 
\begin{align} \label{gaugelag2}
{\mathcal L}_g = \frac 1 \pi adb + {\mathcal L}_M  - j_v  b + \frac 1 \pi  j_q \frac 1 d j_v 
\end{align} 
where the last term denotes that statistical interaction between charges and vortices.  
Also note that by first integrating $\omega$ we get the constraint, $d\tilde b - db = j_q$ which expresses local charge screening.  
Thus \eqref{gaugelag} describes an extreme type II superconductor with vanishing electric screening length, (which is thus {\em not} given by 
$ \lambda_D $), and a finite London length $\lambda_L$. 
The scale $\lambda_D$ does enter in the description of moving vortices, but we do not have any simple understanding of its physical meaning. 

\subsection{Vortices and edges} \label{section:vortedge}
The subgap modes of the p-wave superconductor are bound to the edges and vortices, which, using the lagrangian ${\mathcal L}_g $, have finite extension. In particular, a pointlike static external vortex source, $\rho_{v}=\hbar\delta^{2}(\vec{r})$, generates the fields $B^{a}=\hbar(2\lambda_{L}^{2})^{-1}K_{0}(r/\lambda_{L})$,
$\vec{E}^{b}=-\hbar\alpha_{2}\vec{\nabla}B^{a}$ and $B^{b}=E_{i}^{a}=0$
(in polar coordinates $(r,\theta)$). 
Here $B^{(a)}$ can be interpreted as  the real magnetic field, with spatial extent of $\lambda_L$, penetrating the vortex, 
and $E^{(b)}$ as the associated supercurrent\cite{hos}.

In a chiral p-wave superconductor, the condensate carries an angular momentum proportional to the number of particles \cite{stone04}, which  manifests itself as an edge current. Since the strength of this edge-current $j_q$  thus depends on  the geometry of the system, it is not described by the lagrangian \eqref{gaugelag}, but will be taken as a phenomenological input parameter. 
 Requiring that the gauge fields vanish outside the system, the equations of motion imply that the edge also has a vortex charge density $\rho_v$ proportional to the edge current. For a disc with radius $R$, we can calculate the fields generated by these sources to be $B^b=E^a=0$ and 
\begin{align}
\label{gaugeedge}
B^a(r)&=\frac{\pi \rho_v}{\lambda_L}\frac{I_0(r/\lambda_L)}{I_1(R/\lambda_L)} \\
E^b(r)&=\pi \rho_v\omega^2_p \epsilon_v \frac{I_1(r/\lambda_L)}{I_1(R/\lambda_L)} \ , \nonumber
\end{align}
where $(r,\theta)$ are polar coordinates on the disc, and where $\rho_v$ is a free dimensionful parameter.

\section {The \texorpdfstring{$\gamma$}{gamma-}BF theory} 
\label{section:psiBF}
As already stressed, there are zero modes localized at odd vortices in p-wave paired states. Thus even a topological theory, which should provide   the extreme low-energy description, must include these relevant zero energy modes. To achieve this,
\oncite{hks} proposed the  following minimal extension of the BF theory \eqref{bflag},
\begin{align}
\mathcal{L}_{\gamma BF}= & \frac{1}{\pi}\epsilon^{\mu\nu\rho}\partial_{\mu}a_{\nu}\left(b_{\rho}+\frac{1}{4}\gamma\, i\partial_{\rho}\gamma\right)-j_{{\rm q}}^{\mu}a_{\mu}-j_{{\rm v}}^{\mu}b_{\mu}  \label{tlag}
\end{align}
where  $\gamma$ is a (one-component) Majorana fermion.   The BF term $bda$ is invariant under both parity and time reversal, while the new term $i\gamma d\gamma da$ violates both these symmetries as appropriate for a chiral superconductor. Note that the action for the fermion $\gamma$ has support only where the magnetic field $B^{(a)}=\epsilon^{ij}\partial_{i}a_{j}$ is non zero, and for the bulk theory, this only happens at point like vortices.

To analyze \eqref{tlag}, consider a classical vortex source consisting of $2N$ Wilson lines 
$$
	W_{C_{a}}=\prod_{a=1}^{2N}\exp(im_a\int_{C_{a}}dx^{\mu}b_{\mu})\ ,
$$ where $m_{a}$ are vortex charges, and the curves $C_{a}$ are given
by $x_{a}^{\mu}(t)=\left(t,\vec{r}(t)\right)$. Calculating the corresponding
current, $j_{{\rm v}}^{\mu}(x)$,  substituting in \eqref{tlag}, and integrating out the gauge fields, 
yields the Lagrangian
\begin{align}
L_{{\rm M}}= & \frac{m}{4}\sum_{a=1}^{N}\gamma_{a}(t)i\partial_{\tau}\gamma_{a}(t)\,,\label{partlag}
\end{align}
where we, for simplicity, put all the charges equal to $m$, and where $\gamma_a(t)\equiv\gamma(x_{a}^{\mu}(t))$. 
This describes $N$ gapless Majorana fermions moving along the world
lines of the vortices. In \oncite{hks} the model \eqref{tlag} was analysed further with respect
to fractional statistics, edge modes and ground state degeneracy on the torus.

A problem with the topological field theory \eqref{tlag} is that it predicts zero modes for both odd and even values of the vorticity $m$, while it is known from the BdG description that only odd vortices support zero modes. That this is related to the vortices being pointlike can be understood by considering two very narrowly separated vortices. In the TFT this does not even make sense since there is no length scale - we are simply having two distinct vortices. In the real theory the situation is very different; when the distance between the vortices is comparable with their size, the two Majorana modes can interact and will generically form a gapped Dirac fermion. We thus expect that the TFT \eqref{tlag} emerges in the infrared only for vortices with odd charge. 


\section{The \texorpdfstring{$\psi$}{psi}-BF theory}
We now seek a minimal description of finite size  vortices, and subgap fermions, that, in  a topological scaling limit to be precisely defined later, reduces to  \eqref{tlag} for odd vortex currents.

\subsection{General discussion}
As already pointed out, the potential $a$ is no longer just a statistical gauge field, but describes the actual magnetic field distribution inside the vortex cores. Also, since the charge screening is local we seek a description of the subgap modes in terms of  fermions that are neutral with respect to $a$ but couple to the statistical potential $\omega$. 
Inspired by \oncitec{hks} and anticipating that a pairing interaction will be crucial for obtaining the expected zero modes, we take 
\begin{align}
\mathcal{L}_{\psi}= & \frac{1}{4\pi}\epsilon^{\mu\nu\rho}(\partial_{\mu}a_{\nu})\Psi^{\dagger}( i\partial_\rho + \Gamma_3\omega_\rho)\Psi - \Lambda \Psi^\dagger \Gamma_3 \Psi + \mathcal{L}_\mathrm {pair}     \label{psilag}
\end{align}
where $\Psi^\dagger  = (\psi^\dagger, \psi)$ is  a Nambu spinor, with $\psi$ a dimensionless single component complex fermion field, and where the gamma matrices $\Gamma_i  = \frac 1 2  \sigma_i$ act in the Nambu space.

The first term in \eqref{psilag} is topological and the second, where $\Lambda$ is an energy density, will impose confinement to regions where $B^a$ is non vanishing. Note that this confinement is achieved,  not by introducing a confining potential, but by having the {\em kinetic term for the fermions vanish in bulk}. These two terms have a local U(1) symmetry, $\Psi \rightarrow e^{i \theta\Gamma_3} \Psi$, related to fermion number conservation, and the corresponding gauge field is the statistical potential $\omega$. In a real superconductor, charge and fermion number is only conserved if the superconducting condensate is explicitly taken into account; in a fixed background these quantities are only conserved modulo 2, corresponding to a breakdown of the U(1) symmetry to the discrete symmetry $Z_2$. 
In the BdG description of an s-wave superconductor, this is achieved by a pairing interaction  $\Delta^\star \psi_\uparrow \psi_\downarrow$ which is gauge invariant if $\Delta$ is also transformed under the electromagnetic U(1) gauge group. 
In the case of p-wave pairing, processes corresponding to creation or destruction of Cooper pairs are also associated with absorption or emission of a unit of angular momentum, and the pairing term must be modified accordingly. 

In our effective theory, there is no pairing field, so we must devise another way to describe the breakdown of the U(1) symmetry related to fermion number conservation to a  $Z_2$ symmetry. More precisely, we need a pairing term, ${\mathcal L}_\mathrm{pair}$, that ensures that the lagrangian \eqref{psilag}: 
\begin{itemize}
 \item Has the correct symmetries and is of low order in derivatives. 
  \item Gives rise to a single Majorana zero mode on, in general moving, vortices with odd vorticity. 
\item Supports a fermionic zero mode on the edge of a finite region with an odd bulk vorticity. 
\item Reproduces the known ground state degeneracies on higher genus surfaces. 
\end{itemize}
The crucial step, taken in the two next subsections, will be to express the potential $\omega$ in terms of a co-vector two-frame, $\{e_\mu^a\}$, and then couple the neutral fermions to $\{e_\mu^a\}$ by a pairing term ${\mathcal L}_\mathrm{pair}$. In the following sections we shall then show that with this pairing term the rest of the above conditions are also satisfied. 

\subsection{Field configurations as  a two-frame}
Notice that in the first term in  \eqref{psilag} the space-time derivatives of the fermions are along the direction of 
$f^\mu \equiv  \epsilon^{\mu\nu\alpha} (\partial_\nu a_\alpha )$, and it will be useful to construct a local frame in terms of $f^\mu$ and the two directions orthogonal to it.
Towards this, we first define a metric
\begin{equation}
ds^2 = g_{\alpha\beta} dx^\alpha dx^\beta
=-\frac{\alpha_{2}}{\alpha_{1}}dt\otimes dt+\delta_{ij}dx^{i}\otimes dx^{j} \ ,
\end{equation}
where the factor $\alpha_2 /\alpha_1$ is introduced in order to simplify the
Lagrangian for ${\vec E}^a$, $B^a$. (We could equally well choose 
the ${\vec E}^b $, $B^b$ sector to do this.)
In terms of this metric, the Lagrangian for the gauge fields becomes
\begin{multline}
{\cal L}_M = - \frac{\alpha_2}{8\pi} \, f^a_{\mu\nu}  f^{a \, \mu\nu}
- \frac{\beta_2}{8\pi} f^b_{\mu\nu} f^{b\, \mu\nu} 
\\
+\frac{1}{4\pi} \left( \beta_1 - \frac{\beta_2 \alpha_1}{\alpha_2}\right)
({\vec E}^b)^2
\end{multline}
where $f_{\mu\nu} = \partial_\mu a_\nu - \partial_\nu a_\mu$, and $f^{\mu\nu}=g^{\alpha\mu}g^{\beta\nu} f_{\alpha\beta}$ for both gauge fields. Notice that the choice $\beta_1 = \beta_2\alpha_1 /\alpha_2$ will render this Lagrangian Lorentz invariant. 
From a physics point  of view  this invariance is not required, but it greatly simplifies the calculations of the fermionic spectrum for moving vortices, and will be assumed in the following. 

The construction of an orthonormal frame is now straightforward. We need a set of frame fields
$e^a_\mu$, $a = 0, 1, 2$,  which obey
\begin{align}  \label{ortho}
	 e^a_\mu\, g^{\mu\nu} \, e^b_\nu&=  \eta^{ab} 
\end{align}
where $\eta^{ab} =  \mathrm{diag} (-1, 1, 1)$ is the metric for the tangent frames.
Further, we want to take $e^0$ to be along the direction of
the magnetic field. 
As an explicit realization, we may take
\begin{align}\label{frames}
e^0_\mu &= \frac{1}{\sqrt{- g_{\lambda \sigma} f^\lambda f^\sigma}}\,g_{\mu \alpha} f^\alpha  \, .
\end{align}
The other two frame fields are constructed using \eqref{ortho}, which also means that, by construction, they are orthogonal to $f^\mu$. There is obviously an ambiguity
in solving \eqref{ortho}, since any Lorentz transformation $\Lambda_a^{~k}$ acting on the $e^a$ will give another solution. This is the usual local Lorentz symmetry for frame fields. In our case though, we are fixing $e^0$, so that the only remaining ambiguity is a rotation
${e^a}' = R^a_{~b} e^b$, for $a, b = 1, 2$, where $R^a_{~b}$ is a $2\times 2$ rotation matrix.
As in the case of gravity, this ambiguity of local rotations can be gauged.
For this we introduce a spin connection $\omega_\mu$, but restricted to the
spatial part. One choice for such an $\omega_\mu$ would be to take the standard spin connection which preserves the metric and restrict it to the spatial part.
A simpler choice is to take
\begin{equation}
(\omega_\mu )_{ab} =
\frac12 \left( e^\alpha_a \partial_\mu e_{\alpha \,b}
- e^\alpha_b \partial_\mu e_{\alpha \,a} \right)
\label{omegadef}
\end{equation}
where $a, b = 1, 2$, with $(\omega_\mu)_{0a} = 0$. Thus there is only one component $(\omega_\mu)_{12} \equiv \omega_\mu$. It is easy enough to check that $\omega_\mu$ transforms as a connection under rotations $R^a_{\ b}$,
\begin{equation}
\omega_\mu \rightarrow \omega_\mu + \partial_\mu \Theta
\end{equation}
where $\Theta(x)$ is an angle defined by $R^1_2 = \sin \Theta$. If the rotation $R^a_{\ b}$ of the frame has a non-trivial winding along a closed contour enclosing a vortex, then the field strength $d\omega$ has a flux localized at the vortex.

The notion of how many times a two-frame has rotated along a closed curve is well-defined without any other geometrical structure. What we have done is to use this winding number of the two-frame to encode the information about the vortex flux through any closed loop. The gauge field $\omega$ can then simply be related to the local rotation of the two-frame, which is uniquely defined up to a regular gauge transformation.

The  covariant derivative for fermions with the spin connection is $( \partial_\rho -i  ( \omega_\rho)_{ab} \Gamma^{ab})\Psi$ where $\Gamma^{ab} = -(i/4)[ \Gamma^a , \Gamma^b]$.  Noting that in our case we only need $(\omega_\mu)_{12}$ and that $\Gamma^{12}  = \Gamma^3$, we precisely get the structure appearing in the first, topological, term in \eqref{psilag}. 
It is useful to recall the symmetry properties of this action. First, $L_\psi$ has a full Lorentz invariance under transformations on the world indices $\mu, \, \nu$, etc. (With the choice of $\beta_1 \alpha_2 = \beta_2 \alpha_1$, this applies to the gauge field part of the Lagrangian as well.)
Since we have a flat spac-time,  this is exactly as  expected. The frames $\{e_\mu^a\}$ should be thought of as auxiliary fields, even though their transformation properties may recall many ideas from the theory of gravity. Local rotations on the fields $e^a_\mu$, $a = 1, 2$, is then an additional symmetry. The two first, ``normal", terms in Lagrangian is invariant under the infinitesimal transformation of the fermions given by
\begin{align} \label{efftrans}
\delta \Psi = i \Theta\Gamma_3 \Psi  \\
\delta e^a_\mu = \Theta \epsilon^{ab} e^b_\mu \nonumber \, 
\end{align}
which just expresses that these terms conserve the fermion number.


\subsection{The pairing interaction and  the \texorpdfstring{$\psi BF$}{psi-BF} theory.}

We now construct a pairing interaction which is invariant under the transformation \eqref{efftrans}
and  depends on the background fields $f^\alpha_\mu$ and $e^a_\mu$, which,  of course, are not  
independent because of \eqref{ortho}.
Using that $e^{-i\theta\Gamma_3}\Gamma_a e^{i\theta\Gamma_3} = \Gamma_a + \theta \epsilon^{ab} \Gamma_b + O(\theta^2)$, it is easy to show that 
\begin{align} \label{pair}
	\mathcal{L}_{\mathrm {pair}}  =\frac  {i\delta} {4\pi}\sqrt{\frac{\alpha_2}{\alpha_1}}\epsilon^{\mu\nu\sigma}\, f_\mu \, \Psi^\dagger \Gamma_a e^a_\nu (\partial_\sigma - i \Gamma_3  \omega_\sigma ) \Psi 
\end{align} 
is indeed invariant under \eqref{efftrans}. (The term
$ \omega_\sigma \,\Psi^\dagger \Gamma_a \Gamma_3 \Psi$ is actually zero since it
involves $\psi^\dagger \psi^\dagger$ or $\psi \psi$, and can thus  be omitted.)
When written out, the term \eqref{pair} contains the anomalous operators, $\psi \partial_{\bar z}\psi$ and  $\psi^\dagger \partial_{ z}\psi^\dagger$, where $z = x + iy$ {\em etc}. These operators do not conserve charge, but do conserve the quantity $2f - l$, where $f$ is the fermion number, and $l$ the orbital spin. 
These operators are coupled to the background geometry given by the frame vectors $e^a_\mu$.
It is precisely the connection between charge and orbital spin that makes it possible to eliminate the orderparameter field in favor of a geometric coupling to the frame field, and the electromagnetic field strength, $f^\mu$. Note that for a {\em fixed} background frame field, 
the pairing term \eqref{pair} breaks the U(1) symmetry \eqref{efftrans} down to $Z_2$, just as the BdG Lagrangian with a fixed pairing field 
 $\Delta$.

Having  \eqref{pair}  we can now combine  \eqref{gaugelag} and  \eqref{psilag} to 
\begin{align} \label{fullag}
{\mathcal L}_{\psi BF}=\mathcal{L}_{g}  +\mathcal{L}_{\psi} \, ,
\end{align}
which we shall refer to as the $\psi BF$ theory.


\section{Fermionic subgap modes}
\label{section:subgap}
We now show that the theory defined by \eqref{fullag} indeed supports subgap modes, of which a single one is at zero energy if the vortex is of odd strength. We will consider a static vortex configuration, but since  we have assumed Lorentz invariance, the results in this section equally applies to for the situation of well-separated vorticies moving at a constant velocity.

\subsection{Quantization and vortex configurations}
To quantize the fermions in ${\mathcal L}_\psi$, we shall treat the gauge fields as a classical background, and for a static vortex we get the commutation relations 
\begin{align} \label{eq:comrel}
\left\{ \psi^{\dagger}\left(\vec{r},t\right),\psi\left(\vec{r}^{\,\prime},t\right)\right\} = & \frac{4\pi}{B_a}\delta^{2}\left(\vec{r}-\vec{r}^{\,\prime}\right) \ .
\end{align}

Since in the static case $f^a \propto dt$ equations \eqref{ortho} and \eqref{frames} imply that the two-frame $\{e^a\}$ can be written as
\begin{equation}\label{e-parametrization}
  \begin{aligned}
e^1 &= (\cos(\lambda) \, , \sin(\lambda)) \\
e^2 &= (\sin(\lambda) \, , -\cos(\lambda)) \ ,
  \end{aligned}
\end{equation}
where $\lambda$ is a function that depends on the vortex configuration. Substituting this into \eqref{psilag} yields
\begin{align} \label{BdGlag}
\mathcal{L}_{\psi}= \frac{1}{2\pi} \Psi^\dagger A \Psi
\end{align}
with the matrix,
\begin{align} \label{amatrix}
A = \begin{pmatrix}  i\partial_t - \Lambda &  -\delta B e^{-i\lambda} \partial_{\bar z}   \\
  \delta B e^{i\lambda} \partial_{ z}     & i\partial_t + \Lambda
\end{pmatrix} \ .
\end{align}
 The general solution \eqref{spot}, for vorticies of strength $\{m_a\}$ situated at $\{\vec r_a\}$, and  with $\omega$ substituted with the expression \eqref{omegadef} and $\{e_a\}$ written as \eqref{e-parametrization} is
\begin{align}
	\lambda=\sum_a m_a \arg(\vec r-\vec r_a) +f(t,x,y) \ ,
\end{align}
where $f(t,x,y)$ is a regular function, which can be set to zero by using a gauge transformations of type \eqref{efftrans}.

\subsection{Hamiltonian formulation and boundary conditions for the fermions}
From the Lagrangian \eqref{BdGlag} we get the Hamiltonian
\begin{equation}
\label{fermion-hamiltonian}
	H=\int d^2 x B_a \Psi^\dagger (i\partial_t-A)\Psi \ .
\end{equation}
Quantizing a Lagrangian only gives a formal expression for the Hamiltonian, and to get a well defined time evolution one has to define the domain for this formal expression. In the path integral formulation this amounts to specifying boundary conditions for the operators. Generally, there are many  possible domains for which the Hamiltonian is self adjoint, but in most situations there is only one choice that makes  physical sense. 
For example, there can be an infinite number of ways to make a Hamiltonian self-adjoint using non-local boundary condition involving integrals over space, and where unitarity is preserved by allowing the probability currents to flow between distant points on the boundary. 
Imposing local boundary conditions usually gives a unique self adjoint Hamiltonian, but in our case we shall show that  there are two inequivalent choices.  As a consequence we can have two physically distinct types of edges, but only one them will support a Majorana edge states and thus describe the edge between a topological superconductor and a trivial state.  

A domain for the Hamilonian $H$ specifies the set of operators $\phi$ for which  commutator $[\phi,H]$ is well-defined. This commutator, and thus the time evolution, for operators outside the domain is defined as a limit of a sequence of commutators of operators within the domain,
\begin{equation*}
	i\frac{d}{dt}\phi=\lim_{\alpha\rightarrow 0}\left[\phi_\alpha,H\right] \ .
\end{equation*}

Since the Hamiltonian is quadratic in fermion operators it can be diagonalized by a change of basis, so to obtain the spectrum 
it is sufficient to consider  single particle operators that can be written as,
\begin{equation*}
	\phi_{u,v}^{\dagger}=\frac{1}{4\pi}\int d^{2}xB_{a} \Psi^\dagger(\vec r) \begin{pmatrix} u(\vec{r}) \\ v(\vec{r}) \end{pmatrix}\ .
\end{equation*}
For the operators	 $\phi^\dagger_{u,v}$ to have well defined anti-commutators among themselves, we demand that $(u,v)\in L^{2}(B_{a},M)\times L^{2}(B_{a},M)$, where $M$ is the space-manifold. If $\phi^\dagger_{u,v}$ is in the domain of $[\cdot,H]$ we have 
\begin{equation*}
	[\phi^\dagger_{u,v},H]= \int d^2 x B_a \Psi^\dagger \mathcal H  \begin{pmatrix} u(\vec{r}) \\ v(\vec{r}) \end{pmatrix}
\end{equation*}
with the matrix
\begin{equation} \label{forfig}
	\mathcal H = \frac1B\begin{pmatrix}  \Lambda &  \frac12\delta \{B e^{-i\lambda}, \partial_{\bar z}\}   \\
  -\frac12\delta \{B e^{i\lambda}, \partial_{ z}\}     & -\Lambda
\end{pmatrix} \ .
\end{equation}
Specifying the domain for the operator $\mathcal H$, in the (single particle) Hilbertspace $L^{2}(B_{a},M)\times L^{2}(B_{a},M)$, induces a domain of $[\cdot, H]$, which gives a well defined normal ordered Hamiltonian. For $\mathcal H$ to be self adjoint it has to be symmetric, \emph{i.e.,} $\Braket{\psi|\mathcal H\phi}=\Braket{\mathcal H\psi|\phi}$ for any states $\ket{\psi}$ and $\ket{\phi}$ in the domain of $\mathcal H$. In terms of boundary conditions, this amounts to requiring that the surface term vanishes when we partially integrate, to go from $\Braket{\psi|\mathcal H \phi}$ to $\Braket{\mathcal H\psi|\phi}$. Taking $\phi=(u_1,  v_1)^T$ and $\psi=(u_2,v_2)^T$ we get the boundary term 
\begin{equation*}
	\frac i2\int_{\partial M} \delta B (e^{-i\lambda} v_2^* u_1 dz+e^{i\lambda}u_2^* v_1 d\bar z) \ .
\end{equation*}
where $z$ is the complex coordinate on $M$. A local boundary condition amounts to the requirement that the integrand vanishes. This requirement is solved by
\begin{align}
	 i e^{-i\lambda} u \frac{d z(c(t))}{d t}&=s v \left|\frac{d z(c(t))}{dt}\right| & s&\in \mathbb R \ ,
	\label{boundary-condition}
\end{align}
where $c(t):\mathbb{R}\to \partial M$ is a local parametrization of $\partial M$ in the positive direction induced by the orientation on $M$ given by $\{e^a\}_{a=1,2}$. That is, the positive direction along a boundary of a subset of $\mathbb R^2$ with right (left) handed orientation is given by the direction, which if you followed it, you would have the outward direction to your right (left). 

For the single particle Hamiltonian, $\mathcal H$, each real value of $s$ corresponds to a  self-adjoint extension, but the second quantized Hilbert-space structure gives an additional constraint. Since $[\phi^\dagger,H]^\dagger=-[\phi,H]$ the operator $\phi_{u,v}$ must be in the domain whenever $(\phi_{u,v})^\dagger=\phi_{v^*,u^*}$ is.  This implies that $(v^*,u^*)$ must be in the domain of $\mathcal H$ whenever $(u,v)$ is, and from \eqref{boundary-condition} it then follows that $s^2=1$. 
The remaining sign ambiguity, $s =\pm 1$, is resolved by referring to physics -- as shown below we must take $s=\text{sgn}(\delta B_a/\Lambda)$ to get a Majorana edge mode. This determines the boundary condition \eqref{boundary-condition} and thus, together with the Hamiltonian \eqref{fermion-hamiltonian}, the time evolution.

Since $\mathcal H$ is singular at the vortex insertions, to get a self-adjoint Hamiltonian, 
we have to specify boundary conditions on $u$, and $v$ as they approach these points. 
The simplest way to do this is  to remove a disc or radius $r$ around the vortices at positions $\vec r_a$, and then let $r$ go to zero.
Parametrizing the edge of the disc as  $\vec r=(r \cos(t)+x_a,-r \sin(t)+y_a)$,  
inserting this expression into \eqref{boundary-condition}, and taking the limit $r\rightarrow0$ we get the condition
\begin{equation}
	\label{vortex-bc}
	\lim_{\vec r \rightarrow \vec r_a}e^{i(\arg(\vec r-\vec r_a)-\lambda)}\frac{ u(\vec r)}{v(\vec r)}=s \ .
\end{equation}  

\subsection{Solving the eigenvalue problem}
For the situation of a straight edge and no vortex, or the situation with a single vortex we can analytically find the Majorana mode. Let us first consider the situation with a straight boundary. Without loss of generality we can assume that our system is bounded to the left by the y-axis. By using \eqref{efftrans} we can take $\lambda=0$ and  by taking the limit $R\rightarrow\infty$ of the equations \eqref{gaugeedge} we get the solution $B_a=\pi \rho_v\lambda^{-1}_L e^{-x/\lambda_L}$. Putting this into the expression for $\mathcal H$ the eigenvalue problem 
\begin{equation*}
	\mathcal H \begin{pmatrix} u\\v \end{pmatrix} =E \begin{pmatrix} u\\v \end{pmatrix}
\end{equation*} 
has the solution
\begin{align*}
	(u,v)&=e^{i k y} (\chi(x),\chi^*(x)) & E_k&=\frac{\delta \lambda_L \omega_p^2}{\epsilon_v} k
\end{align*} 
with 
\begin{equation*}
	\chi=\exp\left\{\frac{i\pi}4\left(1-\text{sgn}\left(\frac{\delta B_a}\Lambda\right)\right) +\frac x{2\lambda_L}-\left|\frac{ \lambda_L^2 \Lambda}{\delta \rho_v}\right| e^{x/\lambda_L}\right\}
\end{equation*}
Notice that this solution only is allowed when $s=\text{sgn}(\delta B_a/\Lambda)$ in \eqref{boundary-condition}.

For a single vortex of strength $m$, on the infinite plane, we instead have $B_a=\hbar (2\lambda_L)^{-2}K_0(r/\lambda_L)$, and we can take $\lambda=m\theta$, where $(r,\theta)$ are polar coordinates centered at the vortex. Putting this into the expression for $\mathcal H$ we see that we get the zero energy solution $(u,v)\propto(\chi, \chi^*)$ with 
\begin{multline*}
\chi = \exp\left\{\frac{i\pi}4\left(1-\text{sgn}\left(\frac{\delta B_a}\Lambda\right)\right)\right\}\\
\times\frac1{\sqrt{\left|B_a\right|r}}\exp\left\{i\frac{(m-1)\theta}2-\int^{r}dr\,\left|\frac{\Lambda}{\delta B_a}\right|\right\} \ .
\end{multline*}
Notice that this solution only exists when $m-1\in 2\mathbb Z$ and again only with the boundary condition $s=\text{sgn}(\delta B_a/\Lambda)$.

The above configurations are the only ones we can handle analytically. If we, however, take a rotation invariant system with a vortex in the center of a circular disc, we can  diagonalize the Hamiltonian \eqref{forfig} in  polar coordinates using the ansatz 
\begin{equation}\label{eq:ansatz}
	\begin{pmatrix} u(\vec r)\\v(\vec r) \end{pmatrix} =e^{il \theta} \begin{pmatrix} e^{i(\theta-\lambda)/2}u_l(r)\\e^{-i(\theta-\lambda)/2} v_l(r) \end{pmatrix}\ ,	
\end{equation}
to get the following one-dimensional problem which can be solved using a shooting algorithm:
\begin{align*}
	\partial_r V&=\frac{2\pi}{\delta}\left(\frac{\Lambda}{B_a}+E\right)U-\frac{l}{r}V\\
	\partial_r U&=\frac{2\pi}{\delta}\left(\frac{\Lambda}{B_a}-E\right)V+\frac{l}{r}U\\
	\frac{V(R)}{U(R)}&=1\text{ and } \frac{V(0)}{U(0)}\to -1
\end{align*}
where $U,V $ are $u\sqrt{\frac{rB_a\delta}{4\pi}}$ and $v\sqrt{\frac{rB_a\delta}{4\pi}}$. The parameter $E$ gives the energy eigenvalues when the boundary conditions are satisfied by the solution.

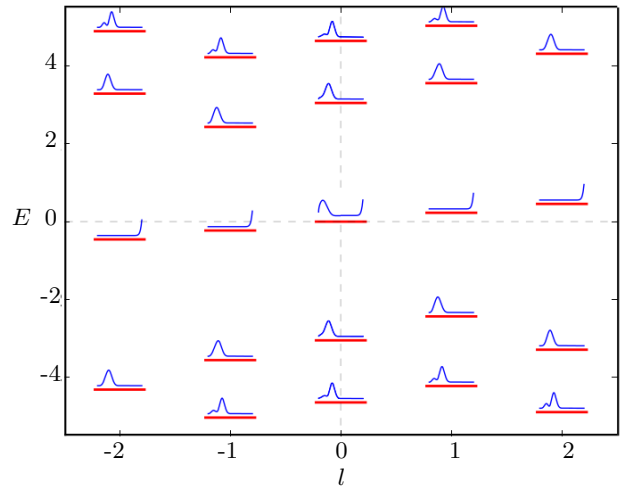
\begin{figure}[!ht]
    \begin{tikzpicture}[x=1.51cm,y=0.52cm]
      \node (plot) at (0,0) {
        \includegraphics[width=0.42\textwidth]{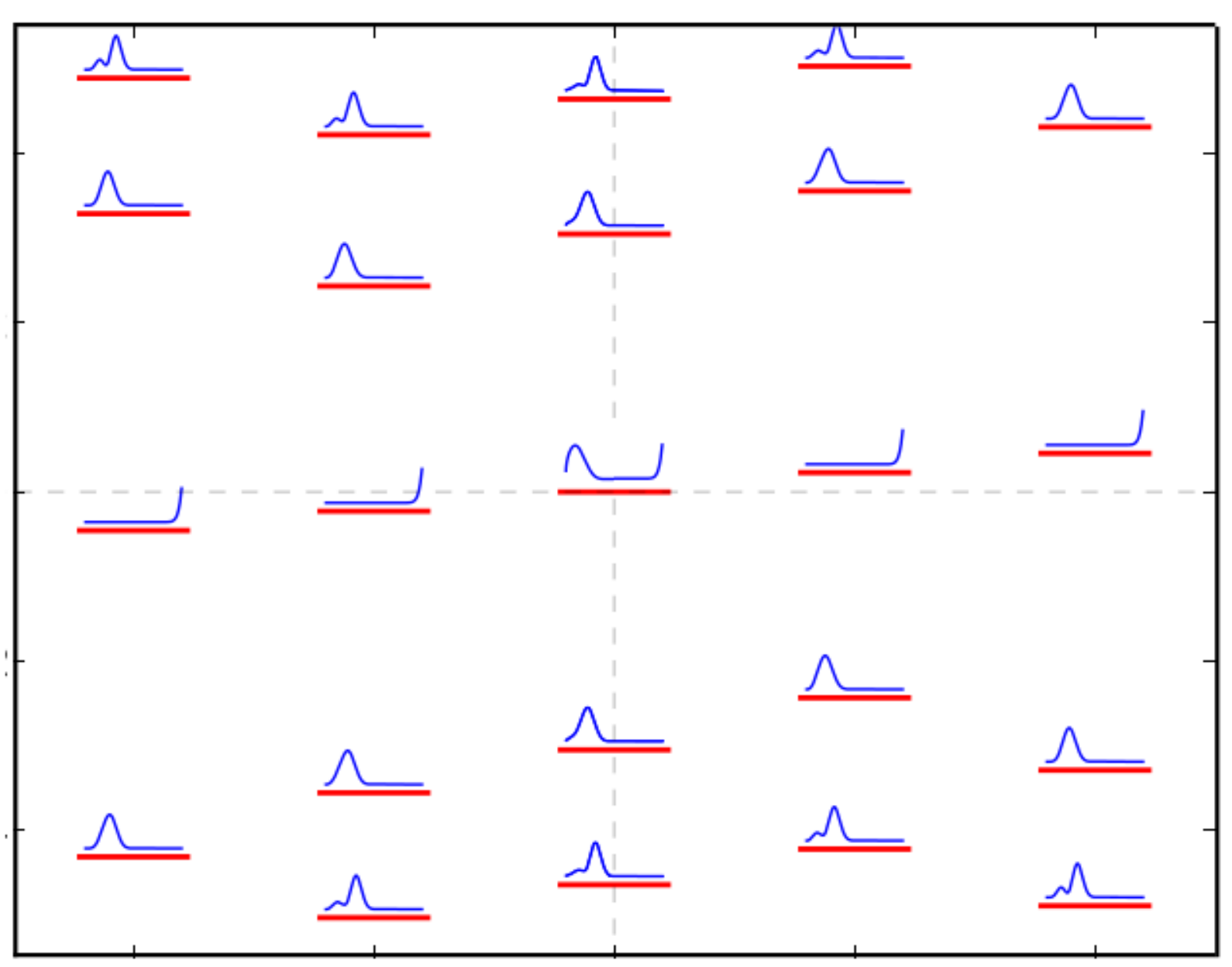}
      };
     \foreach \x in {-2,...,2}  \node[anchor=north] at (\x,-82pt) {\x};
	\node[anchor=north] at (0,-92pt) {$l$};
     \foreach \x in {-4,-2,0,2,4}  \node at (-110pt,\x) {\x};
	\node[anchor=west] at (-128pt,0) {$E$};
    \end{tikzpicture}
    \label{fig:spectrum}
\caption{Energy  as a function of the angular momentum parameter $l$ (see \eqref{eq:ansatz}) for a disc of radius $4.5\lambda_L$. Energy is measured in units of $\Lambda \lambda_L^2$ and the flux, both at the origin and the edge is of unit strength. The small graphs on top of a bar at energy $E$ shows the radial probability density $r\rho(r)=rB_a|\{\Psi^\dagger, \phi_{E}\}|^2$ ($\phi_E$ is the operator with $[H,\phi]=E\phi$) as a function of the distance to the vortex (with increasing distance to the right ). The lowest energy branch is localized at the edge, and the other are localized at the vortex. Note that even though it cannot be seen in the plot there are two (almost) zero energy states at $l=0$.}
\end{figure}

In  Figure \ref{fig:spectrum} we  show the resulting low lying spectrum. The qualitative features are the same as obtained by a self-consistent BdG calculation (see for instance \onlinecite{mizushima10}). In particular, there is a single low energy branch that is localized at the edge, while the remaining modes are higher in energy and are localized at the vortex core. The  gap to the core-excitation is significantly larger than the energy differences  between the core states themselves. 

\section{Ground states  on the torus}
\label{section:GS-degeneracy}
An important characteristic of topologically non-trivial states is the ground state degeneracy on topologically non-trivial manifolds. In the case of the 2D p-wave superconductor this degeneracy is well known, and can be determined using different approaches. Both through the connection to the Pfaffian QH state and the Ising model CFT\cite{read96}, and by a topological classification of the solutions to the BdG equations\cite{read00}, one concludes that on the torus there is a 3-fold degeneracy for fully paired states, while states with an unpaired electron, {\em i.e.} with odd fermion parity, is non degenerate. 

To show that these results also hold true for the effective $\psi$BF theory \eqref{fullag}, we proceed in two steps. First we consider only the gauge part 
\eqref{gaugelag} and show that it has the same four-fold ground state degeneracy as the pure BF theory. Turning to the full $\psi$BF theory we show that the reduction of the number of ground states from four to three, is due to a fermionic zero mode related to a stringlike instanton solution of the Euclidean action in the presence of a topology changing operator. 

Consider a flat torus $(L_x, L_y)$ which admits the topologically nontrivial operators,
\begin{align} \label{orgop}
{\mathcal A}_x(y_0) &= e^{i\oint dx\, a_x(x,y_0)}  \\
{\mathcal B}_y(x_0) &= e^{i\oint dy\, b_y(x_0,y)}  \nonumber
\end{align}
and similarly defined ${\mathcal B}_x(y_0)$ and ${\mathcal A}_y(x_0)$. 
It is understood that these operators are all defined at some fixed time $t$ and  the integrals are taken around a  cycle of the torus. 
These Wilson loop operators have a dual interpretation. For instance, ${\mathcal A}_x$ can either be understood as 
measuring the $a$-flux through the  $x$-cycle of the torus, or as describing a process where 
a quasiparticle quasi hole pair is created and later annihilated after instantaneously encircling the torus in the $x$-direction. 
In a purely topological theory, this concept of an instantaneous process poses no problem, but with a
 Maxwell term present, ${\mathcal A}_x$  should rather be thought of as applying an instantaneous quasiparticle current 
given by $j^q_x(\vec r, t) = L_y^{-1} \delta (y-y_0) \delta (t) $. 
 
 From the BF Lagrangan  \eqref{bflag}, we get the equal time commutation relation
\begin{align} \label{abcom}
[a_i(\vec r_1, t), b_j(\vec r_2,t)] = i \epsilon_{ij}\pi \delta_P^2(\vec r_1 - \vec r_2)
\end{align}
where $\delta_P$ is the periodic delta function on the torus. This leads to the commutation rule
\begin{align} \label{bfalg}
{\mathcal A}_x(y_0) {\mathcal B}_y(x_0) & = e^{-\int d^2r\, [a_x (a_x(x,y_0), b_y(x_0, y)] }{\mathcal B}_y(x_0){\mathcal A}_x(y_0) \nonumber \\
&=  e^{-i\pi} {\mathcal B}_y(x_0){\mathcal A}_x(y_0)  
\end{align}
and similarly  ${\mathcal B}_x(y_0) {\mathcal A}_y(x_0)   +{\mathcal A}_y(x_0)   {\mathcal B}_x(y_0)    =0 $. Note that 
these commutation relations do not depend on the coordinates $x_0$ and $y_0$, so we can obtain 
them also by directly quantizing the quantum {\em mechanical} Lagrangian describing the spatially constant 
modes, which are the only degrees of freedom in the absence of sources\cite{hos}. 

Turning to the Lagrangian ${\mathcal L}_g$, we note that when there are no  sources we can redefine 
$\tilde b \rightarrow \tilde b +  b$ so that the fields  $\omega$ and 
$ \tilde b$ satisfy the same commutation relations as $a$ and $b$ above. Thus, if we define,
\begin{align} \label{modop}
{ \Omega}_x(y_0) &= e^{i\oint dx\, \omega_x(x,y_0)}  \\
\tilde{\mathcal B}_y(x_0) &= e^{i\oint dy\, \tilde b_y(x_0,y)}  \nonumber
\end{align}
we get the same ground state degeneracy as for the pure BF theory. The reader might object to this conclusion, since after the 
shift the fields $(\omega, \tilde b)$ are completely decoupled from $(a, b)$, and naively the latter will give an additional 
fourfold degeneracy. Note, however, that since $a$ does not couple to any source, the corresponding Wilson loop 
does not correspond to any physical process different from that described by ${\Omega }_i$ in \eqref{modop} above. 

In this connection one should note note that things would have been more complicated, had we used the 
Lagrangian ${\mathcal L}_{BF} + {\mathcal L}_M$
considered in \oncitep{hos} This describes a superconductor where both the London and the Debye length are finite, and there is no 
simple commutation relation like \eqref{abcom}. The complication occurs because the application
of any of the Wilson loop operators amounts to an
instantaneous excitation of the system by a quasiparticle or vortex current, and so we will have to identify the resulting new ground state. In 
Appendix A we show how this can be done by introducing topology changing operators that are mildly nonlocal in time. 
As expected, the ground state degeneracy is the same as for the original  ${\mathcal L}_{BF}$ or the Lagrangian ${\mathcal L}_{g}$
describing point like quasiparticles and extended vortices. 

Turning to the full $\psi$BF theory \eqref{fullag}, we first make some remarks concerning its status as a quantum field theory. 
In the previous sections, we avoided questions about the nature of the full quantum ground state, by simply taking a 
classical  configuration for the gauge fields sourced by vortices, and then quantizing the fermions in this background. 
In the absence of a background field there is no kinetic term for the fermions 
(\ie no quadratic term involving one or two time derivatives), and it is not clear how to define a Hamiltonian. We shall thus
define the theory by the path integral, also noting that it is naively, \ie by power counting, renormalizable.\footnote{
This is most easily seen by rescaling the gauge fields to get the canonical mass dimension 1/2, by which the coupling constant
in front of the $a\psi^\dagger \psi$ and $a \psi \psi$ interaction terms also have mass dimension 1/2.} 

We now extract the ground state in the topological sector which is obtained by applying the (unshifted) operator 
$\tilde{\mathcal B}_x(y_0) = e^{i\oint dx\, (\tilde b_x - b_x)(x,y_0))}  $ 
to some eigenstate of $ \Omega_x(y_0)$. As already mentioned, this amounts to introducing a source,
\begin{align} \label{bsourse}
j^v_x(\vec r, t) = L_y^{-1} \delta (y-y_0) \delta (t)
\end{align}
at some fixed time $t$. To extract the ground state wave functional we consider an evolution in imaginary time from $\tau = -\infty$ to $\tau = 0$, with 
the source \eqref{bsourse} inserted at a time $\tau_0 = it_0 \ll 0$, 
\begin{align} \label{grounds}
 {\Psi_0[X]} = \int_{X(-\infty)=X_{in}}^{X(0) = X} {\cal D}[X(\tau)] \, e^{-\int_{-\infty}^0 d\tau \int d^2 r ( {\mathcal L}_{\psi BF} - j_v \tilde b ) }
\end{align} 
where $X = (a,b,\omega,\tilde a,\psi,\psi^\dagger)$ collectively denote all the fields and some suitable boundary conditions,
soon to be discussed, is chosen for the initial state $X_{in}$. We can now treat the source term as a part of the Euclidean action
and evaluate the path integral by first finding the saddle point for the gauge fields. Given the results of the previous sections,
this is  easy. Since the Minkowski action is Lorentz invariant, the Euclidean action is $O(3)$ invariant, and the  solution for a 
static vortex given in section II can immediately be taken over by replacing $r$ with $\sqrt{\tau^2 + y^2}$, if we assume that $L_y \gg 
\lambda_L$ (this condition can be relaxed at the expense of using the full torus Greens functions in terms of theta functions). 
Note that this solution has a finite extent in imaginary time, and is thus best thought of as a stringlike analogue of an instanton. This
analogy is in fact quite apt since, as shown in Appendix X, it changes the eigenvalues of the operator ${\mathcal A}_y$, and thus
connects the different topological sectors.  

The next step is to expand the action in \eqref{grounds} around the saddle point. Had it not been for the fermion part, the wave 
functional could be calculated exactly, since the gauge part is quadratic. Although the original action has no quadratic part in the fermion
action, the background solution provides such a term, just as the static vortex did in section V. To find the spectrum, we must also specify 
the (spatial) boundary conditions on the fermi fields, which can be either periodic  (P) or anti-periodic (A). Here we will  use  
 periodic boundary conditions, and describe the A, or twisted, sectors by a constant vector potential $\omega$. For example,
taking $\omega_x = \pi/L_x$ amounts to having a unit twist in the $x$-direction and the eigenvalue -1 for the operator $\Omega_x$. 

The fermion action in the instanton background is
\begin{align} \label{ferment}
S_x = \int d^3x_E \,  \Psi^\dagger (i\partial_x + \Gamma_3 \omega_x + A) \Psi
\end{align}
where $A$ is the operator given in  in \eqref{amatrix} but with $r= \sqrt{ (\tau - \tau_0)^2 + (y-y_0)^2 }$. 
As shown in Section \ref{section:subgap}, the vector potential $\omega$ can be absorbed by a singular gauge transformation. The same is true here,
with the difference that if $\omega_x$ is an odd multiple of $\pi/L_x$ the boundary condition on the transformed fermions will be 
anti-periodic in the $x$-direction.  Since the eigenvalues of $i\partial_x$ is $2\pi m/L_x$ the action can be written as,
\begin{align} \label{actexp}
S_x = \int d^3x_E\,   \sum_n \left(\lambda_n + \frac {m_n \pi} {L_x} \right) \Psi_n^\dagger  \Psi_n
\end{align}
where the integer $m_n$ is even or odd  for P and A boundary conditions in the $x$-direction respectively, 
and $\lambda_n$ the $n^{th}$ eigenvalues of $A$ as calculated in Section \ref{section:subgap}. 
Thus there is a zero mode in the action when $\omega$ is an even multiple of $\pi/L_x$, \ie for periodic boundary conditions,
and this mode will cause the path integral \eqref{grounds} to vanish. 
We conclude that applying the
operator  $\tilde{\mathcal B}_x(y_0)$ only yields a new state if $\omega_x$ is an odd multiple of $\pi/L_x$ corresponding
to anti-periodic boundary conditions in the $x$-direction. 

Furthermore, if we start from the boundary conditions (P,P) neither $\tilde{\mathcal B}_x$ nor $\tilde{\mathcal B}_y$ will 
yield a new state, so this sector is non degenerate. Starting from any other combination, the two others can be reached 
by applying these operators, so this sector is triply degenerate. For example, starting from (P,A), we can first 
apply $\tilde{\mathcal B}_y$, which, because of the $\Omega_x$ - $\tilde{\mathcal B}_y$ algebra, 
changes the boundary conditions to (A, A), and then apply $\tilde{\mathcal B}_x$ to reach  (A,P). 
Clearly we should now identify the first sector with with states of odd fermion parity, while the second, triply degenerate one,
corresponds to fully paired states with even fermion parity.

\section{The topological scaling limit} 
\label{section:topologicalscaling}
So far, we have shown that the $\psi$BF theory has
all the expected subgap features.
We now show, assuming static (or constant velocity) odd charged vortices, that we retain the $\gamma$BF theory in a well-defined scaling limit. For this we for simplicity assume $\delta B_a/\Lambda>0$, and consider  a collection of $N$ identical vortices of odd strength.
The topological scaling limit is defined by taking
both the physical length scale, $\ell$, and time scale, $\hbar/E$,
to zero at fixed coupling parameters\cite{frohker}.
 We can think of $\ell$ as \emph{e.g.} the minimal distance between the vortices, and
$E$ as a cutoff energy below which our theory is to be valid. 
We define two Majorana fields by, 
\begin{equation}\label{majfield}
  \begin{aligned}
\gamma= & \frac1 2 ( {e^{i(\arg(\vec r-\vec r_a)-\lambda)/2}\psi+e^{-i(\arg(\vec r-\vec r_a)-\lambda)/2}\psi^{\dagger}}) \\
 \tilde{\gamma}= & \frac 1 {2i}({e^{i(\arg(\vec r-\vec r_a)-\lambda)/2}\psi-e^{-i(\arg(\vec r-\vec r_a)-\lambda)/2}\psi^{\dagger}})
  \end{aligned}
\end{equation}
and substitute in \eqref{fullag} to get (setting $j_{q}=0$)  the Lagrangian,
\begin{multline}
\mathcal{L}=\frac{1}{\pi}ad(b+\frac{1}{4}\gamma id\gamma+\frac{1}{4}\tilde{\gamma}id\tilde{\gamma}) +
 \frac 1 \pi d\omega (\tilde b - b) - j_v \tilde b \\
+\frac{1}{8\pi} ad\omega\psi^{\dagger}\psi+{\mathcal L}_M - H \,,\label{eq:subform}
\end{multline}
where $H$ as given by  \eqref{fermion-hamiltonian}. Since $\vec{E}_{a}=0$ we can make the gauge choice $a_0=0$, so the first term in the second line in \eqref{eq:subform} vanishes. Furthermore, the Maxwell terms vanish in the topological limit by the same arguments as in \textcite{hos}.

We now distinguish between having vortices of even or  odd charge. In the first case the spectrum is gapped, so
in the limit where $E$ is taken to be below the subgap, no degree of freedom is left. With odd vortices, the 
 Hamiltonaian $H$ which is of the form $\sum_{E_{n}<E}E_{n}a_{n}^{\dagger}a_{n}$ also vanish in this same limit,
 but the zero modes remain as  degrees of freedom described by the first line in \eqref{eq:subform}. 
To simplify this expression, we make the shift $b\rightarrow b+\frac{1}{4}\tilde{\gamma}id\tilde{\gamma}$, and solve for $\tilde b$
to eliminate the term $ad\tilde{\gamma}id\tilde{\gamma}$ in favor
of $j_{v}\tilde{\gamma}id\tilde{\gamma}$. Using the boundary condition
\eqref{vortex-bc} we see that that $\tilde{\gamma}id\tilde{\gamma}=0$ at vortex cores
so this term in \eqref{eq:subform} also vanishes for a point vortex. This concludes the demonstration
that the topological theory,
\begin{align*}
\mathcal{L}_{\gamma BF}= & \frac{1}{\pi}\epsilon^{\mu\nu\rho}\partial_{\mu}a_{\nu}\left(b_{\rho}+\frac{1}{4}\gamma\, i\partial_{\rho}\gamma\right)-j_{{\rm q}}^{\mu}a_{\mu}-j_{{\rm v}}^{\mu}b_{\mu} \, , 
\end{align*}
proposed in \oncitec{hks} is retained in the scaling limit for odd vortices.

\section{Nonabelian statistics} 
\label{section:non-abelian}
The nonabelian (NA)  statistics in
the Moore-Read QH state was originally understood in terms of the
monodromies in the Ising CFT\cite{NW}, assuming that there are no
remaining Berry phases when the wave functions are represented by
conformal blocks. Proofs for this assertion were given in later papers\cite{read/gurbond}.
In the case of the p-wave superconductor, Ivanov\cite{ivanov} derived
the NA statistics using the BdG formulation of Read and Green\cite{read00}.
Also here it is important that, in a suitably chosen gauge, there
are no Berry phases, so that the braiding phases of the vortices come
entirely from the coupling to the gauge field. Although quite reasonable,
this is not easy to show, and it was taken for granted by Ivanov.
In a later paper\cite{sternetal} Stern \emph{ et al.} addressed this
question, and gave plausible arguments for the absence of Berry phases
by a more detailed analysis of the vortex cores, using certain mild
assumptions about the continuous part of the spectrum. This argument
was later made simpler and more precise by Stone\cite{stone06}.

An important part of Ivanov's argument was that a Majorana operator acquires a minus sign
when encircling another vortex. This, together with a locality assumption, specifies the 
basic braiding operation from which the full action of the braid group can be obtained. 
In \oncite{hks} it was argued that the $\gamma$BF theory also describes NA statistics 
of the Ising type. The argument was quite different from the one given by Ivanov, and used
general properties of Hamiltonian time evolution. The sign in the braid group relations referred to above, 
did not follow from a direct calculation, but was inferred from these general properties. It was conjectured
that the sign would indeed be present in a proper time evolution, but it was pointed out that it was 
difficult to determine due to the singular nature of the action. Above we showed that the proper 
extension of the $\gamma$BF Lagrangian to the $\psi$BF one, both gave a well defined and 
unitary time evolution, and the appropriate minus sign in the braid relation. 
Also, in the $\psi$BF theory there can be no Berry phases, since the fermionic
wave functions only have support on the widely separated vortices.
From this we conclude that the derivation of the NA statistics in \oncite{hks} applies to 
to the $\psi$BF their without using any extra assumptions.  

In \oncitec{hks} the operators corresponding to elementary braid operations were derived by demanding 
certain general properties. It is obviously interesting to try to actually {\em derive} these operators from 
an action. Since the $\psi$BF theory is well defined, this is in fact possible, and we 
shall now derive  the NA statistics by using an alternative
quantization method that directly identifies the Hilbert space for $2N$ vortices as 
a spinor representation of the group $SO(2N)$. This result was originally found
by Nayak and Wilczek for the vortices in the MR pfaffian state. 

First note that for widely separated vortices, moving along the world lines   $\vec{r}_{a}(t) $, the Majorana field  in  \eqref{majfield} 
takes the form, $\gamma(\vec{r},t)=\sum_{a=1}^{2N}\chi \left(\vec{r}-\vec{r}_{a}(t) \right)\gamma_{a}(t)$
where, in an obvious notation, $\gamma_{a}(t)=a_{0,a}(t)+a_{0,a}^{\dagger}(t)$.
Substituting this in the $\psi$BF Lagrangian, taking the topological scaling limit as above, and using  the normalization
of $\chi$, we obtain, once again, the quantum mechanical Lagrangian \eqref{tlag},   
\begin{align*}
L_{{\rm M}}= & \frac14\sum_{a=1}^{2N}\gamma_{a}(t)\, i\, \partial_{t}\gamma_{a}(t)\,,
\end{align*}
where we use the notation, $\gamma_a(t)\equiv\sqrt m \gamma(t,x_{a}^{\mu}(t))$.
The (anti)commutation relations for the corresponding operators $\hat\gamma_i$ follows directly 
from the symplectic structure of the action (for those unfamiliar with this kind of quantization, we 
provide a derivation using standard methods in appendix \ref{appendix:non-abelian}), 
\begin{equation}
\{ \hat\gamma_a ,\hat \gamma_b \} =  2\, \delta_{ab}, \hskip .3in a, \, b = 1, 2, \cdots, 2N
\label{7.6}
\end{equation}
(In the following we shall, for convenience, drop the hat whenever there can be no confusion.) 
In other words, the $\hat\gamma$'s form a $2N$-dimensional Clifford algebra.
This algebra, as is well known, has a unique representation, up to similarity transformations.
The states of the vortices must thus be given by the spinor which carries this representation.

We are interested in the braiding properties of the vortices. For this, consider time-evolution as
given by \eqref{partlag}, starting from an initial set of $\gamma$'s.
 We have the algebra \eqref{7.6} at time $t =0$, and the same algebra is obtained at any later time $t$.
 Thus, time-evolution can at most amount to a similarity transformation of the initial
 $\gamma$'s, so that we can write
 \begin{equation}
 \gamma_a (t) = S^{-1} \, \gamma_a (0) \, S = g_{ab} \, \gamma_\beta(0)
 \label{7.7}
 \end{equation}
where $g_{ab}$ is an $SO(2N)$ rotation.
Thus the braiding properties of the vortices (i.e., any chosen set of initial $\gamma$'s)
can be obtained by using this in the action and quantizing $g_{ab}$.
In terms of the $g_{ab}$, the action is

\begin{equation}
S = - \frac i4\int \left[ ( g^T {\dot g})_{ab} \, \gamma_b(0) \, \gamma_a (0) \right]
\label{7.8}
\end{equation}
Since $g$ is an element of the orthogonal group, $g^T {\dot g}$ is antisymmetric; this is evident from the Grassmann nature of the $\gamma (0)$. 
The matrix with components $i\,\gamma_b(0) \, \gamma_a (0)$ has bosonic (not Grassmann-valued) components and is antisymmetric, so we can regard it as a 
real antisymmetric matrix, the indices $a,\, b$ specifying the matrix elements.
Thus with a $t$-independent orthogonal transformation $\gamma_b \, \gamma_a \rightarrow O^T\gamma_b\gamma_aO$ it can be brought to a quasi-diagonal form,
\begin{equation}
\gamma_b(0) \, \gamma_a (0) =  i\, \sum_{k=1}^N \gamma_{2k-1}(0) \gamma_{2k}(0)~
J_{2k-1~ 2k}
\label{7.9}
\end{equation}
(For more on this issue, see Appendix \ref{appendix:non-abelian}.) In
(\ref{7.9}) $J_{2k-1 ~2k}$ are matrices given by
\begin{equation}
(J_{2k-1 ~2k} )_{ba} = \left\{ \begin{matrix}
-i &\hskip .1in {b = 2k-1, \, a = 2k}\\
i & \hskip .1in {b = 2k, \, a = 2k-1}\\
0&\hskip .1in {\rm other ~values ~of}~b, a\\
\end{matrix} \right.
\label{7.10}
\end{equation}
The $J$'s  correspond to rotation generators in the vector representation; in fact, they are
the generators of the Cartan subalgebra of $SO(2N)$.
Further,  
\begin{equation}
i \gamma_{2k-1} \, \gamma_{2k} = -(1- 2 \, n_k) = \lambda_k \ ,
\label{7.11}
\end{equation}
where $n_k$ counts the occupatian of the fermion $\frac12(\gamma_{2k-1}+i\gamma_{2k})$, 
so the action \eqref{7.8} can finally be written as
\begin{equation}
S = -\frac i4 \int {\rm Tr}  \left[ \lambda_k \, J_{2k-1~2k} ~g^T {\dot g} \right]
\label{7.12}
\end{equation}
Writing the initial state as a superposition of eigenstates of $\{\lambda_k\}$ we can view the time evolution of each of the terms separately, 
and we then have an action only in terms of the bosonic variables $g \in SO(2N)$. 
In  Appendix \ref{appendix:non-abelian} we  apply the method of geometrical quantization\cite{nair06} to this action, 
to show that  a basis for the wave functions, $\Psi(g)$, is given by
\begin{equation} \label{wavefunction}
\Psi_p (g) =  {\cal D}^R_{pw} (g) = \langle 
R_s, p \vert g \vert R_s, w\rangle
\end{equation}
where the state $\vert R_s, w\rangle$ is a highest weight state of weight
$w=( \lambda_1, \lambda_2, \cdots, \lambda_N )$ in the spinorial representation 
of $SO(2N)$ that we denote by  $R_s$; $p$ is a general element in $R_s$.

(The reader familiar with more formal mathematics, will note that 
the action \eqref{7.12} is in the form of the coadjoint orbit action for Lie groups.
The Borel-Weil-Bott theorem (see for instance Ref. \onlinecite{woodhouse97,*perelomov86,*sniatycki80}) 
tells us that the quantization of an action 
\begin{equation}
S = -i \int \sum_{i=1}^r w_i {\rm Tr} \left[ q_i \, g^T{\dot g} \right],
\label{7.13}
\end{equation}
where $g \in G$ and $q_i$ are the generators of the Cartan subalgebra of $G$, will yield 
as a Hilbert space the unitary irreducible representation
of the group $G$ with highest weight $(w_1 , w_2 , \cdots, w_r)$. 
Here $r$ is the rank of the algebra. 
Thus we can immediately conclude that \eqref{7.12} will give a representation
of $SO(2N)$ characterized by the choice of $n_k$. 
From what we said about the representation of the Clifford algebra
\eqref{7.6}, we expect this to be the spinorial representation in \eqref{wavefunction}.)

We can now understand how the braid properties emerge. Starting  with a state given by
$\vert R, w\rangle$ for some choice of the eigenvalues
$\{ \lambda_k \}$. (Recall that we have not fixed the $n_k$ in \eqref{7.11} yet;  there
is still some freedom in the choice of this initial state.)
The state at any time in the future is given by
\begin{equation}
\vert R, p \rangle = 
 {\cal D}^R_{p\,w} (g(t) ) ~\vert R, w\rangle \, .
 \label{7.23}
 \end{equation}
The time-dependence of $g(t)$ is not specified, it can be chosen arbitrarily because
it is a topological theory and observables will be topological in nature.
The ``phase factor" given by time-evolution is nonabelian, being an element
of the spinorial
representation of $SO(2N)$.

Exchanging vortex $1$ and $2$ correspond to a rotation in the
$12$-plane in terms of $SO(2N)$.
The operator corresponding to this is $\theta$
\begin{equation}
\sigma_1 = e^{i J_{12} \theta } \Bigr]_{\theta = \pi/2} =
\frac{1 - \gamma_1 \gamma_2 }{\sqrt{2}}
\label{7.24}
\end{equation}
More generally,
\begin{equation}
\sigma_k = \frac{1 - \gamma_k\, \gamma_{k+1} }{\sqrt{2}}
\label{7.25}
\end{equation}
will exchange the $k$-th vortex with the $(k+1)$-th vortex.
From the properties of the $\gamma$'s we can directly verify that
\begin{align}
\sigma_k \, \sigma_l &= \sigma_l \, \sigma_k , \hskip .3in \vert k - l \vert \geq 2\nonumber\\
\sigma_k \, \sigma_{k+1} \, \sigma_k &= 
\sigma_{k+1}\, \sigma_k \, \sigma_{k+1} 
\label{7.26}
\end{align}
These are the standard braid relations. The phase factors corresponding to the
$\sigma_k$ are given by $ {\cal D}^R_{p\,w} (g )$, $g = \sigma_k$. 
Since these are precisely the nonabelian phase factors corresponding to Ising statistics.

\section{Summary, outlook and comparison to earlier work} 
 In this letter we proposed  the $\psi$BF Lagrangian, equation \eqref{fullag},
as the proper effective theory for the bulk of a  2d spineless p-wave paired superconductor
in the energy range below the minigap. The model also describes the low lying chiral fermion
on the edge of a finite sample, and in particular the zero mode that is present for odd bulk 
vorticity. All known topological properties of the p-wave superconductor is accounted for, 
and in addition the model provides a description of the low-lying spectrum below the 
superconducting gap. 

 Ours is not the first attempt to find a TFT for a p-wave paired state. 
 In a very interesting paper, Fradkin \etal 
 constructed a TFT for the bosonic version of the MR state (which is at $\nu = 1$), 
 using a level 2 non-abelian CS theory, and  showed that by introducing an 
 extra scalar field, one can also describe the original MR state\cite{fradnayak}. 
 Their approach relied on the connection between QH states and conformal field 
 theory (CFT) that was proposed by Moore and Read\cite{mr}. 
 More precisely, the edge theory for the bosonic MR state was shown to be a SU(2) Wess-Zumino-Witten 
 model at level 2, and using the connection between edge and bulk, and the relation 
 between CS theory and CFT, the TFT for the bulk it was identified as a
 non-abelian SU(2) theory at level 2. The theory for the fermionic MR state is more complicated. 
 In addition to the SU(2) gauge field it has a U(1) gauge field and a scalar field, and it was
 argued that it indeed describes a p-wave paired state. We believe that there must be
 a connection between this description, and the one given in this paper, but 
 we have not managed to make it. 
 
 Another challenge is to derive our model from a microscopic theory. This was done in 
 \oncite{hos} for the s-wave case, starting from an abelian Higgs model. To make a similar 
 derivation in the p-wave case is more difficult, since one must find a way to separate 
 and retain the subgap modes (or at the minimum the zero modes) in the effective
 theory.  It is fairly clear how to generalize our model to include s- and d-wave paring, but
 this is unlikely to provide any new insight. The generalization to 3 dimensions is much more
 interesting and here we have already made som limited progress.

\begin{acknowledgments}
We thank M. Sato, A. Stern and X.-L. Qi for interesting discussions at  early stages of this work,  
and E. Fradkin, A. Karlhede and S. Ryu for helpful
discussions and comments on the manuscript. THH is supported by the
Swedish Research Council, and VPN's work was supported by the U.S.\ National
Science Foundation grant PHY-1213380 and by a PSC-CUNY award.
\end{acknowledgments}

\appendix  
\section  {Ground state degeneracy in the BF-Maxwell theory}  \label{appendix:GS-degeneracy}

In this appendix we show how to properly define an operator ${\mathcal B}_i$ in the BF-Maxwell (BFM) theory so that it has the same
 topological properties as the corresponding operator in the pure BF theory.  For simplicity we shall use the Lorentz invariant version of the BFM action 
 (but we feel confident that our results will apply also to systems without this invariance) which in Feynman gauge reads,
 \begin{multline}
{\mathcal L}_{BFM}=  \frac{1}{\pi}  \epsilon^{\mu\rho\nu}a_\mu\partial_\rho b_\nu +  \frac 1 {2\pi\mu} a_\mu  g^{\mu\nu}\Box a_\nu \\
+  \frac 1 {2\pi\mu} b_\mu g^{\mu\nu}\Box b_\nu  - j_q a - j_v b
\label{bfm2lag}
\end{multline}
where $\mu$ is the topological mass well known from Chern-Simons Maxwell theory\cite{deser82}.  
Introducing $A^T = (a_\mu, b_\mu)$, and $J^T = (j_q, j_v)$,  \eqref{bfm2lag} can be expressed as
\begin{align} \label{short lag}
{\mathcal L}_{BFM}= \frac 1 2 A^T G^{-1} A - \frac 1 2  J^T A - \frac 1 2 A^T J \, .
\end{align}
Inverting the kernel $G^{-1}$ gives the propagator,
\begin{align} \label{matprop}
G = \frac {\pi \mu^2} {\Box + \mu^2} \left(\begin{array}{cc}\frac 1 \mu & -\frac d \Box \\ -\frac d \Box & \frac 1 \mu\end{array}\right)  
+ \mathrm {gauge \ terms}
\end{align}

We shall only need  the off-diagonal elements
\begin{align} \label{abcomp}
G_{ab} = G_{ba} = - \frac {\pi\mu^2} {\Box + \mu^2} \frac d \Box \, ,
\end{align}
and in particular its $xy$ component, which in momentum space is,
\begin{align} \label{momspg}
G^{xy}_{ab} (\omega, \vec p) = - \frac {\pi\mu^2} {\omega^2 - p^2 - \mu^2 + i\epsilon} \frac {i\omega} {\omega^2  - p^2 + i\epsilon} 
\end{align}

We now show how the canonical equal time commutator \eqref{abcom} for the 
 in the pure BF theory can be derived from the Lagrangian using the method introduced by Johnson and 
 Low\cite{johnson66} in the context of current algebra. For this we first note that the pure BF theory is obtained 
 from the BFM theory by taking the limit $\mu \rightarrow \infty$. 
Applying the method from \oncitec{johnson66} and using the large $\mu$ limit of \eqref{abcomp},  the equal time commutator is  given by,
\begin{align} \label{jltrick}
&[a_x(\vec r_1,0), b_y(\vec r_2,0] = \nonumber \\ 
&= \lim_{\tau \rightarrow 0} [ i G^{xy}_{ab}(\vec r_1 - \vec r_2, \tau) - i  G^{xy}_{ab}(\vec r_1 - \vec r_2, -\tau)  ] \\
&= \lim_{\tau \rightarrow 0} \int \frac {d^2p} {(2\pi)^2}\int \frac { d\omega}{2\pi} \frac {\pi \omega(e^{i\tau\omega} - 
e^{-i\tau\omega} )} { \omega^2 - p^2  + i\epsilon} e^{i \vec p\cdot (\vec r_1 - \vec r_2)} \nonumber
\end{align} 
The $\omega$ integral in the two terms can  be closed upwards and downwards in the complex plane respectively, 
resulting in a closed clock-wise contour, $C_\infty$, at infinity,
\begin{align*} 
\oint_{C_\infty}  \frac {d\omega} {2\pi} \frac \omega {\omega^2 - p^2 } = i \, .
\end{align*}
Finally the $d^2p$ integral give a delta function to reproduce the result \eqref{abcom} obtained by canonical quantization. 

The corresponding $\omega$-integral for the full BFM theory is,
\begin{align} \label{fullexp}
\int \frac { d\omega}{2\pi}  \frac {-\pi\mu^2} {\omega^2 - p^2 - \mu^2 + i\epsilon} \frac {\omega(e^{i\tau\omega} - 
e^{-i\tau\omega} )} {\omega^2 - p^2  + i\epsilon} 
\end{align}
which is zero since the integrand falls as $\omega^3$ at infinity. A heuristic way to understand this is to 
realize that the equal time commutator is related to  the large $\omega$ part of the propagator, which is due to the 
usual Maxwell term which has no ab-component. 

It is rather clear what is going wrong. Because of the Maxwell terms, the world lines of the particles and the vortices 
get an thickness $\sim 1/\mu$, and we only expect to get a well defined braiding phase when the lines are kept a distance apart that 
exceeds this thickness. This means that we must take $\mu \tau \gg 1$ and evaluating the two terms
${\mathcal A}_x(y_0) {\mathcal B}_y(x_0)$ and ${\mathcal B}_y(x_0){\mathcal A}_x(y_0)$ separately. Since the 
theory is quadratic, the calculation is trivial.
Defining 
\begin{align} \label{fullcurr}
J^\mu(\tau) = \left(\begin{array}{c} \oint dx\, \delta(y-y_0)\delta(t-\tau/2), 0,0 \\ 0,  \oint dy\, \delta(x-x_0)\delta(t+\tau/2),  0\end{array}\right)
\end{align}
we immediately get
\begin{align} \label{expo}
\bra 0 {\mathcal  A}_x (y_0,\tau/2) {\mathcal B}_y (x_0,-\tau/2) \ket 0 &= e^{-\frac 1 2 J^T(\tau) \hat G J(\tau) } \\
\bra 0 {\mathcal B}_y (x_0,\tau/2) {\mathcal  A}_x (y_0, -\tau/2)  \ket 0 &= e^{-\frac 1 2 J^T(-\tau) \hat G J(-\tau) } \nonumber
\end{align}
where $\hat G$ is the torus version of \eqref{matprop}. Explicitly we have, using  translational invariance, 
\begin{align}
 J^T(\tau) \hat G J(\tau)  &=  \oint dx_1\, \oint dx_2 \, \hat G_{aa}^{xx} (x_1 - x_2, 0,0) \nonumber  \\ 
 &+ \oint dy_1\, \oint dy_2 \,\hat  G_{bb}^{yy} (0, y_1 - y_2, 0,0)  \\
 &+  2\oint dx\, \oint dy \, \hat G_{ab}^{xy}(x,y,\tau)  \, . \nonumber
 \end{align} 
 The two first terms will give contributions proportional to the length of the strings, that in a fully Lorentz invariant theory would be 
 just the radiatively generated masses of  the charges and point-vortices respectively. This contribution will be the same for the
 two lines in \eqref{expo}, so to find the relation between them we only need to evaluate the last integral, and using that the $\vec p =0$ component for the torus propagator, $\hat G$, is the same as the one in infinite space, $G$, (This can be 
derived using the relation $\hat G(x,y,t) = \sum_{m,n = -\infty}^\infty G(x+m L_x, y+n L_y, \tau)$ where $L_x$ and $L_y$ are the 
lengths of the cycles of the torus.) we get, 
 \begin{align}
 \hat G(\vec p = 0,\tau) &=  
 \int\frac {d\omega} {2\pi}  \frac {\pi\mu^2} {\omega^2  - \mu^2 + i\epsilon} \frac {-i\omega e^{i\omega\tau}} {\omega^2  + i\epsilon} \\
&= - \frac \pi 2 (1+e^{-i\mu\tau})
\end{align}
where we closed the contour in the upper half plane to get the last identity. Substituting in \eqref{expo} gives,
\begin{align} \label{result1}
& \bra 0 {\mathcal  A}_x (y_0,\tau/2) {\mathcal B}_y (x_0,-\tau/2) \ket 0 \\
&+ e^{i\pi \cos\mu\tau} \bra 0 {\mathcal B}_y (x_0,\tau/2) {\mathcal  A}_x (y_0, -\tau/2)  \ket 0 = 0 \nonumber \, .
\end{align}
We notice that for $\tau = 0$ the two operators commute, which is consistent with the vanishing of the integral
\eqref{fullexp}. We also understand what is needed to regain the anticommutator characteristic of the pure BF theory;
the operators should be smoothened so that the oscillating factor $\cos\mu\tau$ cancels. This is easily obtained by 
redefining one or both of the operators. Since we working in a limit of extended vortices, and pointlike charges, 
it is natural to define,
\begin{align} \label{redefinition}
\tilde {\mathcal B}_y(x_0,t) &= e^{i\int_{-\infty}^\infty dt' \, f(t'-t) \oint dy\, b_y(x_0,y)} 
\end{align}
where $f(t)$ is a function peaked at $t=0$ and a widths $\Delta  \ll 1/\mu$. With this definition,  \eqref{result1}
is consistent with the BF commutation relation \eqref{bfalg}. The result \eqref{result1}
for the vacuum expectation value, is in fact true for any matrix element of the operators. 
The easiest way to see this is to work in a basis of coherent in and out-states of the $a$ and $b$ fields,
\begin{align} \label{cohstate}
	\ket {\vec \alpha_i(\vec r) }= N e^{  \vec \alpha(\vec r) \cdot \vec  a^\dagger_{\mathrm in} (\vec r) } \ket 0
\end{align}
where $\vec  a_{\mathrm in} (\vec r)$ is the in-field operator, and similarly for the out-filelds, and the 
$\vec  b_{\mathrm in} (\vec r)$, and $\vec  b_{\mathrm out} (\vec r)$. $N$ is a normalization constant that 
can be determined, but which is not needed for the argument.
With these states it is  straightforward to evaluate the general matrix elements 
$
\bra {\vec \alpha_{\mathrm out}, \vec\beta_{\mathrm out}} {\mathcal  A}_x (y_0,\tau/2) {\mathcal B}_y (x_0,-\tau/2) 
\ket {\vec \alpha_{\mathrm in}, \vec\beta_ {\mathrm in} }
$
{\em etc.},  and conclude that the extra contributions, compared to those in \eqref{expo} are the same for the two 
ordering of the operators ${\mathcal A}_x$ and $\tilde {\mathcal B}_y$. Thus we have shown that these two operators
 anti-commute as long as they are evaluated with a time separation $\tau \gg \Delta\gg 1/\mu$. This is clearly all that is needed to 
conclude that the degeneracy of the BFM theory is the same as for the pure BF theory. 


\section  {Quantiztion of the Majorana Lagrangian }  \label{appendix:non-abelian}

There are three issues we will discuss in this appendix. First, we show how the commutation rules (\ref{7.6}) arise, then we carry out the geometric quantization of
the action (\ref{7.9}) and finally we conclude with some clarifying remarks
on the quasi-diagonalization of $i\, \gamma_b (0) \gamma_a (0)$ in the passage from
(\ref{7.8}) to (\ref{7.12}).

To derive \eqref{7.6} is useful to write the action in a slightly more transparent form by defining the 
usual complex fermions $b_k$ and $b^*_k$, $k = 1, 2, \cdots, N$ by
\begin{equation}
\gamma_{2k-1} = (b_k + b^*_k ), \hskip .3in 
\gamma_{2k} = -i \, ( b_k - b^*_k )
\label{7.2}
\end{equation}
In terms of these variables, the action for \eqref{partlag} becomes
\begin{equation}
S = i\, \sum_{k=1}^N \,\int dt~ b^*_k \, {\dot b_k}
\label{7.3}
\end{equation}
where we have removed a total derivative. The canonical one-form is given by the boundary value for the time-integration in $\delta S$. This is easily seen to be
$i \sum b^*_k \,\delta b_k$
where $\delta $ is to be interpreted as the exterior derivative on the space of 
the variables $b^*_k, \, b_k$.
If we take the wave functions to be a function of the $b$'s, which is a fermionic coherent state
description, it directly follows that,
\begin{equation}
b^*_k \, \Psi = \frac{\partial \Psi}{\partial b_k} \, .
\label{7.4}
\end{equation}
Keeping in mind that the
variables $b$, $b^*$ are Grassmann-valued, this equation corresponds to the anticommutation rule
\begin{equation}
b_k \, b^*_l + b^*_l \, b_k = \delta_{kl}
\label{7.5}
\end{equation}
which, using \eqref{7.2} yields \eqref{7.6} in the text.


Next we turn to the derivation
of the expression  \eqref{wavefunction} for the basis functions in the Hilbert space corresponding
to 2N vortices.
The canonical structure for the action \eqref{7.12} is given by
\begin{equation}
\Omega = \frac i4 {\rm Tr} ( \lambda_k  J_{2k-1~2k} ~ g^{-1} \delta g \wedge g^{-1} \delta g ) \, .
\label{7.14}
\end{equation}
This is seen to be invariant under $g(t) \rightarrow g(t) \, h(t)$, $h = \exp( i J_{2k-1~2k} \varphi_k(t) )$, 
where $\varphi_k(t)$ are time-dependent angles describing the rotation, so 
that $\Omega$ is defined on $G/H$, where $H$ is Cartan subgroup.
This means that $g \rightarrow g \, h$ is a gauge transformation, and  that the physical  phase space is $G/H$.

The simplest approach to quantization is via coherent states.
For this we begin by  considering wave functions $\Psi (g)$,
which are functions of $g$, \ie  they are defined on the full phase space. We then 
impose a holomorphicity condition, as is appropriate for coherent states.
(In the language of geometric quantization, this amounts to choosing a polarization 
of the prequantum  wave functions.)
For this, consider the (right) group translations on $g$ given by
\begin{equation}
R_A \, g = g \, t_A
\label{7.15}
\end{equation}
where $\{ t_A \}$ is an orthonormal basis for the Lie algebra of $SO(2N)$.
In the Cartan basis, we can group them as the raising operators
$R_{+i}$, $i = 1, 2, \cdots, N (N-1)$, the lowering operators
$R_{-i}$ and the generators of the Cartan subalgebra
$R_a$, $a = 1, 2, \cdots, N$. (The last set corresponds to 
$\{ J_{2k-1 ~2k}\}$ for the case of the vector representation.)
The holomorphicity condition can be taken as
\begin{equation}
R_{+i} \, \Psi (g) = 0 \, .
\label{7.16}
\end{equation}

As the next step, we consider
 the right transformation of $g$ in \eqref{7.12} as
$g \rightarrow g \, h$, $h = \exp ( i J_{12}\, \varphi )$, \emph{i.e.} a gauge transformation. Since 
$J_{12}$ commutes with all $J_{2k-1~2k}$, we find from \eqref{7.12}
\begin{equation}
S \rightarrow S + \int dt\,  \frac{\lambda_1}2\, {\dot \varphi} \, ,
\label{7.17}
\end{equation}
and since wave functions transform as $e^{i S}$, we find 
\begin{equation}
\Psi (g \, h ) = \Psi ( g ) ~ \exp \left( i \frac{\lambda_1}2  \varphi \right) \, .
\label{7.18}
\end{equation}
More generally, we get
\begin{equation}
\Psi (g \, h) =  \Psi (g) ~ \exp \left( i \sum \frac{\lambda_a}2  \varphi_a \right) \, .
\label{7.19}
\end{equation}
The solution to the conditions \eqref{7.16} and \eqref{7.19} is obtained as follows.
The Peter-Weyl theorem tells us that, as functions on $G$, $\Psi (g)$  can always be written as 
\begin{equation}
\Psi (g) = \sum_{R,p,q} C^R_{pq} \, {\cal D}^R_{pq} (g) = \sum C^R_{pq} \, \langle 
R, p \vert g \vert R, q\rangle
\label{7.20}
\end{equation}
where $ {\cal D}^R_{pq} (g)$ is the matrix representation of $g$ in the unitary irreducible
representation $R$, $p, \, q$ being the matrix labels.
The holomorphicity condition \eqref{7.16} means that the state $\vert R, q\rangle$ must be a highest
weight state. Further, the condition \eqref{7.19} shows that the eigenvalues of the Cartan
subalgebra correspond to $( \lambda_1, \lambda_2, \cdots, \lambda_N )$.
In other words, a basis for the wave functions obtained by quantization of \eqref{7.12} is
given by
\begin{equation}
\Psi_p (g) =  {\cal D}^R_{pw} (g) = \langle 
R, p \vert g \vert R, w\rangle
\label{7.21}
\end{equation}
where the state $\vert R, w\rangle$ is a highest weight state of weight
$( \lambda_1, \lambda_2, \cdots, \lambda_N )$. 

Going back to the definition of $J$'s we see that
\begin{equation}
[J_{12}, \gamma_1 ] = -i \gamma_2 , \hskip .1in [ J_{12}, \gamma_2 ] = i \gamma_1, \hskip .2in
{\rm etc.}
\label{7.22}
\end{equation}
Thus the spinor version of the $J$'s is given by
$J_{2k-1 ~2k} = i \gamma_{2k-1} \gamma_{2k} /2$. This identifies $\lambda_k$ as the eigenvalues
of the Cartan elements in the spinor representation.
Using this, the representation $R$ for ${\cal D}^R_{pw} (g) = \langle 
R, p \vert g \vert R, w\rangle$ in \eqref{7.21} is identified as the spinor representation.
This completes the explicit argument for constructing the wave functions.


Finally, we turn to some clarifying remarks about the passage from (\ref{7.8}) to (\ref{7.12}).
The point to worry about is that the entries of $B$ are bilinears of Grassmann-valued variables.
In the theory of quantization of coadjoint orbit actions, such cases have, to our knowledge,  not been considered,
which is why we felt that further clarification was needed.

We start by writing the action (\ref{7.8}) as
\begin{equation}
S = - \frac i4\int \left[ ( g^T {\dot g})_{ab} \, B_{ba} \right] =  - \frac i4\int \mathrm{Tr} [ g^T {\dot g}\, B]
\label{7.8a}
\end{equation}
where 
\begin{equation}
B_{ba} = \gamma_b(0) \, \gamma_a (0) 
\label{7.8b}
\end{equation}
Since $g$ is an element of the orthogonal group, $g^T {\dot g}$ is antisymmetric (up to boundary terms), and so is $B$ as is evident from the Grassmann nature of the $\gamma (0)$.
Notice that only the antisymmetric part of $B$ will contribute to the trace anyway,
even if we consider $B$ as just a matrix made of commuting variables.

To define the dynamics, we need an initial value for the real, bosonic, $2N\times 2N$ matrix $B$. This amounts to assigning real numbers to the entries in the matrix and we show that this can be done by specifying only $N$ real numbers. First assume that an arbitrary assignment 
of real numbers as the entries of $B$ has been made, \ie
$B \rightarrow B^{in}$, where the entries of $B^{in}$ are arbitrary real numbers. We can now put $B^{in}$ in  block-diagonal form by an orthogonal transformation 
\begin{equation} \label{transf}
B^{in} = O\, B^{in;J} \,O^T
\end{equation}
with
\begin{equation*}
B^ {in;J} =  i\, \sum_{k=1}^N \lambda_k  J_{2k-1~ 2k}
\end{equation*}
where  $\lambda_k = 2n_k - 1$ are real numbers, and $J_{2k-1 ~2k}$ are matrices given by
(\ref{7.10}). So far, there is no reason why the values
of $n_k$ should be only $0$ and $1$ for fermions, or  non-negative integers for bosons.
Substituting \eqref{transf} in the action \eqref{7.8} and using the cyclicity of the trace and the fact
that $B^{in}$ is 
time-independent and relabeling $gO \rightarrow g$,  we get, 
\begin{equation}  
S = - \frac i4\int \mathrm{Tr} [ g^T {\dot g}\, B^{in;J}]     \, .
\label{action2}
\end{equation}
This is basically the result (\ref{7.12}) apart from the question of the values
of $n_k$.

We now argue that a  consistent quantization  will constrain the allowed values of 
$\lambda_k$, and thus $n_k$  to those appropriate for bosons or fermions. 
For this consider again the transformation
$g \rightarrow g\,h $, $h = \exp( i J_12 \varphi (t))$, where we take $g$ to be independent of time
(this can be easily generalized without any change in the conclusion), and with $\varphi(t) $ changing by $2\pi$ over the entire
time-evolution. Thus $h$ goes from the identity to the identity over the time-evolution.
The wave function changes by $\exp (i \lambda_1 \pi )$ from
(\ref{7.18}). Since we are back at the same point on the group,
we must have $\exp (i \lambda_1 \pi ) =1$ for the bosonic case, or $\exp (i \lambda_1 \pi ) = -1$
if we allow double-valued wave functions as for the fermionic case.
In case of fermions, the  consistency will require quantization of the entries in $B^{in;J}$ to be
half integers, corresponding to integer values of $n_k$, just as for bosons.

We must still explain why, in the fermionic case the values of $n_k$ is restricted to 0 and 1. 
The action for a general quantum system 
with an initial state specified by a density matrix $\rho_0$, can be taken as,
\begin{equation}
S = \int_0^t  dt ~ {\rm Tr} \left[ \rho_0 \, U^\dagger \left( i \frac{\partial U}{ \partial t} - H \, U\right) \right] \, ,
\label{action3}
\end{equation}
since the equation of motion obtained by varying
$U$ is the standard quantum Liouville equation for the density matrix
$\rho (t) = U \, \rho_0 \, U^\dagger$.
In our case, the time evolution is also a symmetry transformation corresponding to a group 
$G$, which means that $U$ is a representation of $G$ and can be written as 
\begin{equation}
U = \exp \left( i {\hat T}_a \theta^a \right)
\label{action4}
\end{equation}
where ${\hat T}_a$ are the operators on the quantum Hilbert space 
 corresponding to the symmetry generators $t_a$ of the group.
We then have, just by group property,
\begin{equation}
i\, U^\dagger \frac{\partial U }{ \partial t}  = {\hat T}_a \, E^a_{~b} \, d\theta^b
\label{action5}
\end{equation}
where $E^a_{~b}$ are functions of $\theta$ defined by
\begin{equation}
i\, g^{-1} \frac{\partial g }{ \partial t}  = t_a \, E^a_{~b} \, d\theta^b
\label{action6}
\end{equation}
Using (\ref{action5}) in (\ref{action3}) we see that the kinetic  term involving the time-derivative
is
\begin{equation}
S_k = \int dt \, \langle {\hat T}_a \rangle \,E^a_{~b} \, d\theta^b 
= -\frac{i}{ 4}  \int dt\, {\rm Tr} \left(  g^{-1} {\dot g} \, B\right) 
\label{action7}
\end{equation}
where $\langle {\hat T}_a \rangle =
{\rm Tr}( \rho_0 {\hat T}_a )$ and $B =
- 8 \langle {\hat T}_a \rangle \, t_a$, with the normalization
${\rm Tr} (t_a t_b ) = \frac{1}{ 2} \delta_{ab}$.
Thus the action is of the form
(\ref{7.8}) or (\ref{7.8a}) and that the matrix $B$ is related to the expectation value
of the quantum operators in the initial state.
Since we already established that the Hilbert space at any time is that of $N$ Dirac fermions, it follows
by consistency, that the number operators have to be assigned the values 0 or 1, 
even though $\gamma_b (0) \gamma_a (0)$ in (\ref{7.8b}) involves Grassmann variables. 
This concedes the discussion about the route from (\ref{7.8}) to (\ref{7.12}).

Finally note that while for the bosonic case, this discussion might be a bit of an overkill, it does
provide an argument for why the $n_k$ have to be taken as \emph{positive} integers, since that does
not follow from the argument based on gauge invariance only.

\bibliography{pwave}
\bibliographystyle{apsrev4-1}

\end{document}